\documentclass[a4paper,11pt]{article}
\pdfoutput=1 % if your are submitting a pdflatex (i.e. if you have
\usepackage{amsmath}                    % Matemáticas, poner separado del resto (incompatibilidad de no sé qué)
\usepackage[bbgreekl]{mathbbol}
\DeclareSymbolFontAlphabet{\mathbbl}{bbold}
\usepackage{amsfonts,amssymb,amsthm,mathtools}    % Para escribir en Matemáticas
\DeclareSymbolFontAlphabet{\mathbbm}{bbold}
\DeclareSymbolFontAlphabet{\mathbb}{AMSb}%
\usepackage{tensor,mathrsfs}
\usepackage[utf8]{inputenc} %utf8                    % Para mejorar la bibliografía
\usepackage[T1]{fontenc}
\usepackage{geometry} 				
\usepackage{lmodern}
\usepackage{enumerate}
\usepackage{titlesec}
\usepackage{bm}
\usepackage[dvipsnames]{xcolor}            % Nos permite el manejo de las opciones de color.
\usepackage[customcolors]{hf-tikz}
\usepackage{array}
\usepackage{authblk}
\usepackage[numbers,sort&compress]{natbib}
\usepackage{scalerel}
\usepackage{tensor}
\usepackage{jheppub}
\usepackage{accents}
\usepackage{todonotes}

\graphicspath{{./font/}{./}}

\newcommand{\dd}{\mathchoice
	{\mathbbm{d}\rrule{.087ex}{1.605ex}\hspace*{0.15ex}} % display
	{\mathbbm{d}\rrule{.087ex}{1.605ex}\hspace*{0.15ex}} % inline
	{\mathbbm{d}\rrule{.08ex}{1.125ex}\hspace*{0.15ex}}  % sub/super script
	{\mathbbm{d}\rrule{.06ex}{.8ex}\hspace*{0.15ex}}     % subsub/supersuper script
}

\newlength{\alturaL}
\newcommand{\LL}{\includegraphics[height=1.1\alturaL]{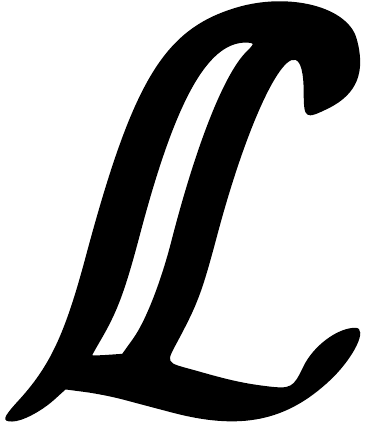}}

\newlength{\alturaO}
\newcommand{\OOmega}{\includegraphics[height=\alturaO]{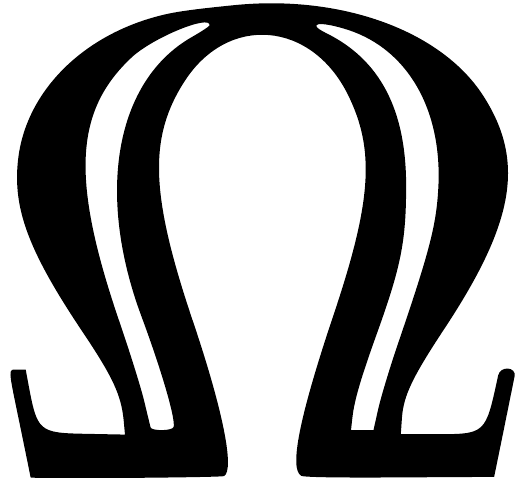}}

\newlength{\alturaI}
\newcommand{\ii}{\raisebox{-.04ex}{\includegraphics[height=1.1\alturaI]{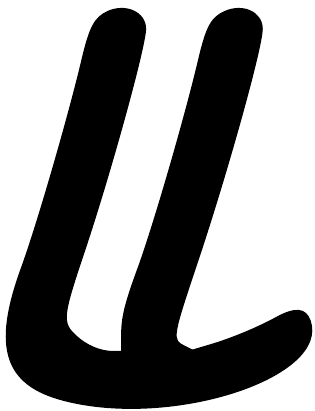}}}

\newlength{\alturaJ}
\newcommand{\jj}{\raisebox{-.37ex}{\includegraphics[height=.9\alturaJ]{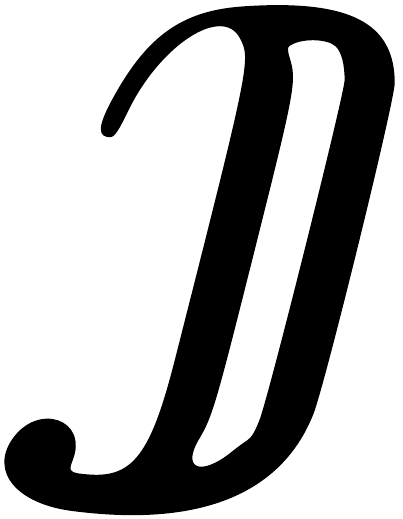}}}

\newlength{\alturaX}
\newcommand{\campos}{\raisebox{-.075ex}{\includegraphics[height=1.05\alturaX]{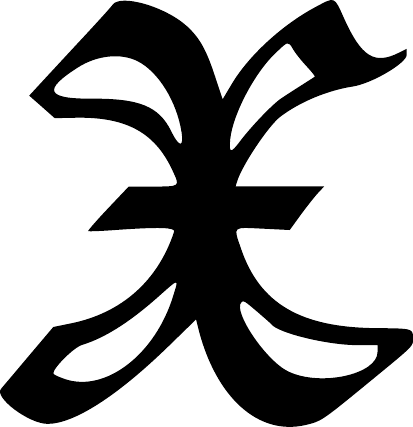}}}

\newcommand{\comentario}{\hfsetfillcolor{green!5}\hfsetbordercolor{green!50!black}}
\renewcommand{\thesection}{\Roman{section}} %%%%%%%%

\usepackage{xpatch}
\makeatletter
\AtBeginDocument{\xpatchcmd{\@thm}{\thm@headpunct{.}}{\thm@headpunct{}}{}{}}
\makeatother

\usepackage[nottoc]{tocbibind}          % Añade la bibliografía, lista de tablas, imágenes...
\usepackage{pstricks}
\usepackage{tikz}

\usetikzlibrary{cd}
\usepackage[title]{appendix}

\usepackage{mathtools}
\usepackage{enumitem}
\usepackage{verbatim}

\makeatletter
\newsavebox\myboxA
\newsavebox\myboxB
\newlength\mylenA

\newcommand*\xoverline[2][0.65]{%
    \sbox{\myboxA}{$\m@th#2$}%
    \setbox\myboxB\null% Phantom box
    \ht\myboxB=\ht\myboxA%
    \dp\myboxB=\dp\myboxA%
    \wd\myboxB=#1\wd\myboxA% Scale phantom
    \sbox\myboxB{$\m@th\overline{\copy\myboxB}$}%  Overlined phantom
    \setlength\mylenA{\the\wd\myboxA}%   calc width diff
    \addtolength\mylenA{-\the\wd\myboxB}%
    \ifdim\wd\myboxB<\wd\myboxA%
       \rlap{\hskip 0.5\mylenA\usebox\myboxB}{\usebox\myboxA}%
    \else
        \hskip -0.5\mylenA\rlap{\usebox\myboxA}{\hskip 0.5\mylenA\usebox\myboxB}%
    \fi}
\makeatother

\newcommand{\corurl}{BrickRed}  \newcommand{\corcite}{red}
\newcommand{\corlink}{blue}    \newcommand{\corfile}{black}

\hypersetup{
linktocpage,
colorlinks,
urlcolor=\corurl,	
citecolor=\corcite,
linkcolor=\corlink,
filecolor=\corfile,
pdfnewwindow,
pdftitle={Covariant space methods GR},
pdfauthor={Juan Margalef Bentabol},
pdfsubject={CPS}}

\allowdisplaybreaks  
\makeatletter
\newcommand{\pushright}[1]{\ifmeasuring@#1\else\omit\hfill$\displaystyle#1$\fi\ignorespaces}
\newcommand{\pushleft}[1]{\ifmeasuring@#1\else\omit$\displaystyle#1$\hfill\fi\ignorespaces}
\makeatother

\newcommand{\updown}[3]{\overset{#1}{\underset{#2}{#3}}}   % \updown{regla}{cadena}{=}

\makeatletter
\newcommand{\raisemath}[1]{\mathpalette{\raisem@th{#1}}}
\newcommand{\raisem@th}[3]{\raisebox{#1}{$#2#3$}}
\makeatother
\newcommand{\peqsub}[2]{#1_{\raisemath{.2pt}{\hspace*{-0.1ex}\scriptscriptstyle #2}\hspace*{-0.2ex}}}

\makeatletter
\DeclareRobustCommand{\rvdots}{%
	\vbox{
		\baselineskip4\p@\lineskiplimit\z@
		\kern-\p@
		\hbox{.}\hbox{.}\hbox{.}
}}
\makeatother

\newcommand{\Q}{\mathbb{Q}}     % Racionales
\newcommand{\R}{\mathbb{R}}     % Reales
\renewcommand{\S}{\mathbb{S}}   % Esfera
\def\QED{{\boldmath$\rule{0.5em}{0.5em}$}}                                % Definimos cómo es el QED (final de la demostración)
\def\markatright#1{\leavevmode\unskip\nobreak\quad\hspace*{\fill}{#1}}    % Definimos: colocar a la derecha ¿no?
\def\qed{\markatright{\QED}}                                              % Definimos el qed final (QED colocado a la derecha

\newtheorem{theorem}{Theorem}[section]
\newcommand{\lateral}{\peqsub{\partial}{L}}

\makeatletter
\let\c@equation\c@theorem
\makeatother
\usepackage{chngcntr}
\counterwithin{equation}{section}
\newtheorem{definition}[theorem]{Definition}       
                   
\newtheorem{remark}[theorem]{Remark}

\newtheorem*{lemma*}{Lemma}

\theoremstyle{definition}

\usepackage{graphicx}
\makeatletter
\newcommand\rrule[3][0pt]{%
	\ifdim#2>#3\math@hrule[#1]{#2}{#3}\else\math@vrule[#1]{#2}{#3}\fi}
\newcommand\math@hrule[3][0pt]{%
	\gdef\mystery@factor{0.07}%
	\@tempdima=#3%
	\rule[#1]{0pt}{#3}% Needed to account for .5\@tempdima vertical offset of rounded rule
	\raisebox{.5\@tempdima+#1}{%
		\makebox[#2][l]{\kern-.5\@tempdima\@@mathrule{#2}{#3}}}%
}
\newcommand\math@vrule[3][0pt]{%
	\gdef\mystery@factor{0.0}%
	\@tempdima=#2%
	\rule[#1]{0pt}{#3}% Needed to account for .5\@tempdima vertical offset of rounded rule
	\raisebox{-.0\@tempdima+#1}{%
		\kern0.5\@tempdima%
		\rotatebox{90}{\kern-0.5\@tempdima\makebox[#3][l]{\@@mathrule{#3}{#2}}}%
		\kern0.5\@tempdima}%
}
\def\@@mathrule#1#2{%
	\@tempdimb=#2%
	\@tempdima=\dimexpr#1-\mystery@factor\@tempdimb%Why 0.07 for \math@hrule?
	\pdfliteral{%
		q []0 d %
		1 J %  set line cap to rounded ends
		\strip@pt\@tempdimb\space w \strip@pt\@tempdimb\space 0 m %
		\strip@pt\@tempdima\space 0 l S Q }}
\makeatother

\makeatletter
\DeclareFontFamily{OMX}{MnSymbolE}{}
\DeclareSymbolFont{MnLargeSymbols}{OMX}{MnSymbolE}{m}{n}
\SetSymbolFont{MnLargeSymbols}{bold}{OMX}{MnSymbolE}{b}{n}
\DeclareFontShape{OMX}{MnSymbolE}{m}{n}{
<-6>  MnSymbolE5
<6-7>  MnSymbolE6
<7-8>  MnSymbolE7
<8-9>  MnSymbolE8
<9-10> MnSymbolE9
<10-12> MnSymbolE10
<12->   MnSymbolE12
}{}
\DeclareFontShape{OMX}{MnSymbolE}{b}{n}{
<-6>  MnSymbolE-Bold5
<6-7>  MnSymbolE-Bold6
<7-8>  MnSymbolE-Bold7
<8-9>  MnSymbolE-Bold8
<9-10> MnSymbolE-Bold9
<10-12> MnSymbolE-Bold10
<12->   MnSymbolE-Bold12
}{}

\let\llangle\@undefined
\let\rrangle\@undefined
\DeclareMathDelimiter{\llangle}{\mathopen}%
{MnLargeSymbols}{'164}{MnLargeSymbols}{'164}
\DeclareMathDelimiter{\rrangle}{\mathclose}%
{MnLargeSymbols}{'171}{MnLargeSymbols}{'171}
\makeatother

\newcommand{\F}{\mathcal{F}}

\renewcommand{\d}{\mathrm{d}}

\renewcommand{\SS}{\mathbb{S}}
\renewcommand{\L}{\mathcal{L}}
\newcommand{\XX}{\mathbb{X}}
\newcommand{\QQ}{\mathbb{Q}}

\newcommand{\Sol}{\mathrm{Sol}}

\newcommand{\Lag}{\mathrm{Lag}}

\newcommand{\vol}{\mathrm{vol}}

\def\equivintt{{\setbox0\hbox{\ensuremath{\mathrel{\equiv}}}\rlap{\hbox to \wd0{\hss\ensuremath\int\hss}}\box0}}
	
\newcommand{\equivint}{\mathrel{\equivintt}}

\def\wwedgee{{\setbox0\hbox{\ensuremath{\mathrel{\wedge}}}\rlap{\hbox to \wd0{\hss\,\ensuremath\wedge\hss}}\box0}}

\newcommand{\wwedge}{\mathrel{\!\wwedgee\!}}

\def\ledgee{{\setbox0\hbox{\ensuremath{\mathrel{\cdot}}}\rlap{\hbox to \wd0{\hss\ensuremath\wedge\hss}}\box0}}

\newcommand{\ledge}{\mathrel{\ledgee}}

\def\uptolidss{{\setbox0\hbox{\ensuremath{\mathrel{=}}}\rlap{\hbox to \wd0{\hss\raisebox{1.2ex}{\ensuremath{\scriptscriptstyle\mathcal{L}}}\hss}}\box0}}

\newcommand{\uptolids}{\mathrel{\uptolidss}}

\title{Covariant phase space for gravity with boundaries: metric vs tetrad formulations}

\author[c,d]{J. Fernando Barbero G.}   
\author[a,c]{Juan Margalef-Bentabol}
\author[b,c]{Valle Varo}
\author[b,c]{Eduardo J.S.~Villaseñor}

\emailAdd{fbarbero@iem.cfmac.csic.es}
\emailAdd{juanmargalef@psu.edu}
\emailAdd{valle@cvb.es}
\emailAdd{ejsanche@math.uc3m.es}

\affiliation[a]{Institute for Gravitation and the Cosmos and Physics Department. Penn State
	University. PA 16802, USA.
	\vspace*{1ex} \mbox{}}
\affiliation[b]{Departamento de Matemáticas, Universidad Carlos III de Madrid. Avda. de la
	Universidad 30, 28911 Leganés, Spain.
	\vspace*{1ex} \mbox{}}
\affiliation[c]{Grupo de Teorías de Campos y Física Estadística. Instituto Gregorio Millán (UC3M).
	Unidad Asociada al Instituto de Estructura de la Materia, CSIC, Madrid, Spain.}
\affiliation[d]{Instituto de Estructura de la Materia, CSIC, Serrano 123, 28006, Madrid, Spain}

\abstract{We use covariant phase space methods to study the metric and tetrad formulations of General Relativity in a manifold with boundary and compare the results obtained in both approaches. Proving their equivalence has been a long-lasting problem that we solve here by using the cohomological approach provided by the relative bicomplex framework. This setting provides a clean and ambiguity-free way to describe the solution spaces and associated symplectic structures. We also compute several relevant charges in both schemes and show that they are equivalent, as expected.}

\arxivnumber{}

\begin{document}

\maketitle

\flushbottom

\section{Introduction}

Space-time boundaries play a prominent role in classical and quantum General Relativity (GR). Their applications range from  black hole thermodynamics \cite{Wald,WaldLR} to the study of radiative modes at \textit{scri} \cite{P1,P2,AS,ACL}. They are also fundamental to describe isolated and dynamical horizons \cite{AK,AshtekarFairhurstKrishnan2000,AshtekarEngleSloan2008,EKPP10,EKPP11} or in the definition of physical charges \cite{IyerWald,CorichiReyes2015,BrianDolan,MargalefVillasenor2020}. Boundary terms also control the dynamical properties of some midisuperspace models with ``conical singularities'' at spatial infinity such as Einstein-Rosen waves coupled to massless scalar fields \cite{AM,ALarge,BGV,BV}.\vspace*{2ex}

Classical GR defined in a space-time without boundary may be alternatively described by the first order Hilbert-Palatini (or Holst) action in terms of tetrads or the Einstein-Hilbert Lagrangian in metric variables. In particular, the corresponding Hamiltonian descriptions are equivalent. Moreover, by employing standard covariant phase space (CPS) methods, it is possible to see that their respective symplectic potentials differ by an exact form. Thus, when this form is integrated over a Cauchy slice $\Sigma$, the result vanishes as a consequence of Stokes' theorem. This shows that the presymplectic forms corresponding to the metric and the tetrad formalisms are equivalent in the space of solutions.

The situation changes substantially when boundaries (possible at conformal infinity) are present. Some of the concerns regarding the equivalence of the tetrad and metric formulations may be attributed to the choice of bulk and boundary Lagrangians. In fact, many discrepancies found and discussed in the literature \cite{JacobsonMohd2015,OliveriSpeziale2020a,PaoliSpeziale2018,FGP1} are to be expected since the boundary dynamics is not always explicit and the solution spaces are not fully specified. To avoid these problems, it is crucial to choose variational principles such that the equivalence of the dynamics they describe, both in the bulk and at the boundaries, is guaranteed. The choice of the bulk and boundary Langrangians will determine the presymplectic structure in the solution space as well as the Noether charges.\vspace*{2ex}
 
The purpose of this paper is to prove that the metric and tetrad formulations are equivalent in the CPS with Dirichlet boundary conditions (BC) and (homogeneous) Neumann BC (the same methods can be applied to other BC). We do this by relying on the relative bicomplex framework, a formalism developed in \cite{MargalefVillasenor2020} which is cohomological in nature so no \emph{ad hoc} choices are required. This formalism will allow us to establish the equality of the metric and tetrad symplectic potentials in the \emph{relative cohomology} (equality up to a \emph{relative exact form} as explained in the appendix). Consequently, the respective presymplectic forms in the solution space are equivalent and, as expected, there is a precise correspondence between the Noether charges in both formulations. As a direct result of our discussion, we will consider the asymptotically flat case and rederive the well-known formula for the ADM-energy.\vspace*{2ex}

The paper is structured as follows. In the next section, we present some of the geometric concepts relevant for the implementation of the CPS-algorithm developed in \cite{MargalefVillasenor2020}.  In \ref{section: metric}, we apply this algorithm to the metric formulation of GR. Section \ref{section: tetrads} presents an analogous study for the tetrad formalism. In Section \ref{section: metricvstetrad} both formulations are compared and shown to be equivalent. We present our conclusions in the last section of the paper. We have also included in appendix \ref{section: appendix} a short summary of the relative bicomplex framework together with some computational details.

\section{The geometric arena}

\subsection[\texorpdfstring{$M$}{M} as a space-time]{\texorpdfstring{$\boldsymbol{M}$}{M} as a space-time}\label{subsection: M space-time}
Let $M$ be a connected and oriented  $n$-manifold admitting a foliation by Cauchy hypersurfaces. Without loss of generality, $M=I\times \Sigma$ for some interval $I=[t_i,t_f]$, ($t_i<t_f$) and some $(n-1)$-manifold $\Sigma$ with boundary $\partial\Sigma$ (possibly empty). Denoting $\Sigma_i=\{t_i\}\times\Sigma$ and $\Sigma_{\!f}=\{t_f\}\times\Sigma$, we split $\partial M$ into three distinguished parts
\[\partial M=\Sigma_i\cup \peqsub{\partial}{L} M\cup \Sigma_{\!f}\]
where $\Sigma_{i,f}$ are the ``lids'' and $\lateral M:=I\times \partial \Sigma$ the ``lateral boundary''. $M=I \times \Sigma$ is a manifold with corners $\partial\Sigma_i\cup\partial\Sigma_{\!f}$, which, as a set are $\partial(\lateral M)$. The following diagram summarizes relevant notational information about embeddings and the induced geometric objects.

\begin{equation}\label{eq: diagrama}
\begin{tikzcd}[row sep = scriptsize]
\Big(\Sigma,\gamma,D,\{a,b,\ldots\}\Big)\arrow[rr,"{(\imath,n^\alpha)}",hook]  & &
\Big(M,g,\nabla,\{\alpha,\beta,\ldots\}\Big)\\
\Big(\partial\Sigma,\xoverline{\gamma},\xoverline{D},\{\xoverline{a},\xoverline{b},\ldots\}\Big) 
\arrow[u, "{(\xoverline{\jmath},\mu^a)}",hook]
\arrow[rr,"{(\xoverline{\imath},\xoverline{m}^{\overline{\alpha}})}"', hook] & &
\Big(\partial M,\xoverline{g},\xoverline{\nabla},\{\xoverline{\alpha},\xoverline{\beta},\ldots\}\Big)
\arrow[u,"{(\jmath,\nu^\alpha)}"', hook]
\end{tikzcd}
\end{equation}
The entries in each $4$-tuple in (\ref{eq: diagrama}) are: the manifold, the (non-degenerate) metric or pulled-back metric, its associated Levi-Civita connection, and the abstract indices used to describe tensors in the manifold. The arrow labels specify the notation used for embeddings and normal unitary vector fields. Horizontal arrows are associated with future pointing normal unit vector fields and vertical arrows are used for outward pointing normal unit vector fields at the boundary. The notation just introduced is not consistent for the bottom lid $\Sigma_i$, as it may be thought of as a spacelike hypersurface embedded by $\imath_i$ or considered as part of $\partial M$\!. $\Sigma_i$ often appears as the boundary of $M$\!, so we choose the outwards (past pointing) convention in this case. An overline is often used to denote objects that live exclusively at the boundary, such as $\xoverline{g}$, which in index notation reads $\xoverline{g}_{\overline{\alpha}\overline{\beta}}$. Notice that the embeddings $\jmath$ and $\xoverline{\jmath}$ are fixed since the boundaries are also fixed.  
However, $\imath$ and $\xoverline{\imath}$, which embed $(\Sigma,\partial\Sigma)$ into $(M,\partial M)$, can be chosen among the Cauchy embeddings satisfying $\imath(\partial\Sigma)\subset\lateral M$\!. \vspace*{2ex}

As $(M,g)$ is oriented, we have the metric volume form $\vol_g$ that assigns the value $1$ to every positive orthonormal basis. We orient $\Sigma$ and $\lateral M$ with $\vol_{\gamma}$ and $\vol_{\overline{g}}$, respectively given by
\begin{align}\label{eq: orientation sigma}
\imath^*(\iota_{\vec{U}}\vol_g)=-n_\alpha U^\alpha\vol_\gamma\qquad\qquad\qquad \jmath^*(\iota_{\vec{U}}\vol_g)=\nu_\alpha U^\alpha\vol_{\overline{g}}
\end{align}
for every $\vec{U}\in\mathfrak{X}(M)$. These orientations are the ones for which Stokes' theorem holds in its usual form. Finally, $\partial\Sigma$ can be oriented as the boundary of $\Sigma$. Thus $\vol_{\overline{\gamma}}$ is given by
\begin{align}\label{eq: reverse orientation}
\xoverline{\jmath}^*(\iota_{\vec{V}}\vol_\gamma)=\mu_a V^a\vol_{\overline{\gamma}}\qquad\quad\longrightarrow\quad\qquad\xoverline{\imath}^*(\iota_{\vec{W}}\vol_{\overline{g}})=+\xoverline{m}_{\overline{\alpha}} W^{\overline{\alpha}}\vol_{\overline{\gamma}}
\end{align}
Notice that if we use the Stokes' theorem from $\lateral M$ to $\partial(\lateral M)=\partial\Sigma_i\cup\partial\Sigma_{\!f}$, a minus sign appears in the integral over $\partial\Sigma_i$.

\subsection[From \texorpdfstring{$M$}{M}-vector fields to \texorpdfstring{$\F$}{F}-vector fields]{From \texorpdfstring{$\boldsymbol{M}$}{M}-vector fields to \texorpdfstring{$\boldsymbol{\F}$}{F}-vector fields}
Let $\F$ be  a space of tensor fields on $M$ (sections of a certain bundle $E\stackrel{\pi}{\to}M$) and think of it as an infinite dimensional differential manifold where we assume that the usual differential objects---like tensor fields, the exterior derivative $\dd$, the interior derivative $\ii$, the wedge $\wwedge$\,, or the Lie derivative $\LL$---are well defined (see the appendix of \cite{MargalefVillasenor2020} for more details about this and other technicalities). As a reminder for the reader, most objects defined over the space of fields are denoted with a double font. Given a vector field $\xi\in\mathfrak{X}(M)$ on $M$,  let us associate to it a canonical vector field $\XX_\xi\in\campos(\F)$ on $\F$. To that purpose, we assume that $\xi$ is tangent to $\lateral M$ but not necessarily to $\Sigma_i$ and $\Sigma_{\!f}$ (notice that we can always extend the interval $[t_i,t_f]$ to avoid problems at the lids) and take advantage of the fact that a field $\phi^r$ ($r$ ranging over the number of fields of the theory) can be interpreted in two ways:
\begin{itemize}
	\item As a tensor field on $M$\!. That is, a section $\phi^r:M\to E^r$ of some bundle $E^r\overset{\pi}{\to}M$ such that $\phi^r_p:=\phi^r(p)\in E^r_p:=\pi^{-1}(p)$. In particular, we can take its Lie derivative $(\L_\xi\phi^r)_p=\partial_\tau|_{0}(\varphi^\xi_\tau)^*\phi^r_p$, where $\{\varphi^\xi_\tau\}_\tau\subset\mathrm{Diff}(M)$ is the flow of $\xi$. If $\F$ is reasonable enough, as it is the case in this paper, we will have $\L_\xi\phi\in\F$.
	\item As a point of $\F$. A vector field $\XX_\xi\in\campos(\F)$ is a section of the tangent bundle $T\F$ i.e.\ $(\XX_\xi)_\phi:=\XX_\xi(\phi)\in T_\phi\F\cong\F$. The last isomorphism comes from the fact that the field space $\F$ is linear in this case. The non-linear case is not as straightforward but, in concrete examples, it is possible to carry out similar constructions.
\end{itemize}
Taking into account these remarks, we define
\begin{equation}\label{eq: definicion XX_xi}
(\XX_\xi^r)_{\phi}=\L_\xi\phi^r\qquad\qquad\equiv\qquad\qquad\LL_{\XX_\xi}\phi^r=\L_\xi\phi^r
\end{equation}
Notice that $\dd\phi^r$ (which appears in the definition of the Lie derivative $\LL$ through Cartan's magic formula) has to be interpreted as the exterior derivative of the evaluation function.
	
\subsection{CPS-Algorithm}\label{section: summary of algorithm}
This paper relies heavily on the results of \cite{MargalefVillasenor2020}.  For convenience, a summary has been included in the appendix but we list here the main steps of the CPS-algorithm  since it will be used throughout the paper. This algorithm provides an ambiguity-free method to construct a presymplectic structure  in the space of solutions canonically associated with the action of the theory.

\begin{enumerate}\setcounter{enumi}{-1}
	\item Given the action $\SS:\F\to\R$ describing the dynamics of a particular field theory, choose any Lagrangian pair $(L,\xoverline{\ell})$ such that
	\[\SS=\int_M L-\int_{\partial M}\xoverline{\ell}\]
	\item Compute $\dd L=E_r\wedge\dd\phi^r+\d\Theta$ and choose any $\Theta$ compatible with the previous expression (uniquely defined up to a $\d$-exact form).
	\item Compute $\dd\xoverline{\ell}-\jmath^*\Theta=\xoverline{b}_r\wedge\dd\phi^r-\d\xoverline{\theta}$  over $\lateral M$\! and choose any admissible $\xoverline{\theta}$.
	\item Define $\Sol(\SS)=\{\phi\in\F\ /\ E_r(\phi)=0, \ \ \xoverline{b}_r(\phi)=0\}$ and the inclusion $\peqsub{\jj}{\SS}:\Sol(\SS)\hookrightarrow\F$.
	\item Compute the presymplectic structure canonically associated with $\S$
	\[\OOmega_\SS^\imath:=\dd\!\left(\int_\Sigma\imath^*\Theta-\int_{\partial\Sigma}\overline\imath^*\xoverline{\theta}\right)\qquad\longrightarrow\qquad\OOmega_\SS=\peqsub{\jj}{\SS}^*\OOmega^\imath_\SS\]
	where $\imath:\Sigma\hookrightarrow M$ is any Cauchy embedding and $\xoverline{\imath}:=\imath|_{\partial\Sigma}:\partial\Sigma\hookrightarrow\lateral M$ its restriction. $\OOmega_\SS$ does not depend on $\imath$.
\end{enumerate}
It is important to notice that it is not always possible to perform the second step. If $\xoverline{b}_r$ and $\xoverline{\theta}$ do not exist (their existence does not depend on the chosen $\Theta$), that means that the theory is not well defined and hence we will have to change $\F$ and/or the action (hence $(L,\xoverline{\ell})$) to ensure that the equality holds. Once we have the presymplectic structure, the following two steps provide important additional information about the theory at hand.

\begin{enumerate}\setcounter{enumi}{4}
	\item Study symmetries; i.e. find out if $\XX_\xi$ is a $\underline{\d}$-symmetry and obtain the $\xi$-currents and $\xi$-charges.
	\[\begin{array}{l}
	J_\xi:=\iota_\xi L-\ii_{\XX_\xi}\Theta\\
	\xoverline{\jmath}_{\xi}:=-\iota_{\overline{\xi}}\xoverline{\ell}-\ii_{\XX_\xi}\xoverline{\theta}
	\end{array}\qquad\qquad\qquad \QQ^{\imath}_\xi:=\int_\Sigma\imath^*J_\xi-\int_{\partial\Sigma}\xoverline{\imath}^*\xoverline{\jmath}_{\xi}\]
	\item Compare $\OOmega_\SS$ with the presymplectic structure obtained in the  standard Hamiltonian formulation.
\end{enumerate}

\section{General Relativity in terms of metrics}\label{section: metric}

\subsection*{Step 0: Action}
Let us consider the following spaces of metrics on $M$
\begin{align*}
&\F^{(m)}_N=\mathrm{Met}(M)&&\F^{(m)}_D=\left\{g\in\mathrm{Met}(M)\ /\ \xoverline{g}:=\jmath^*\!g\text{ fixed}\right\}
\end{align*}
that we will refer to as Neumann and Dirichlet metrics respectively. Here $\mathrm{Met}(M)$ is the space of Lorentzian metrics on $M$ such that the lateral boundary is time-like and the lids are space-like (it is an open set of the space of all Lorentzian metrics on $M$). Consider also the actions $\SS^{(m)}_N$ and $\SS^{(m)}_D$ both equal to
\[\SS^{(m)}(g)\!=\!\!\int_M\! L^{(m)}_{\mathrm{EH}}(g)-\int_{\partial_L\!M}\!\xoverline{\ell}^{(m)}_{\mathrm{GHY}}(g)\qquad\qquad\begin{array}{l}
L^{(m)}_{\mathrm{EH}}(g)=(R_g-2\Lambda)\vol_g\\[1.2ex]\xoverline{\ell}^{(m)}_{\mathrm{GHY}}(g)=-2 \mathrm{Tr}_{\overline{g}}(\xoverline{K})\vol_{\overline{g}}
\end{array}\]
but suitably restricted to these spaces of metrics. Our Lagrangian pair $(L^{(m)}_{\mathrm{EH}},\xoverline{\ell}^{(m)}_{\mathrm{GHY}})$ consists of  the Einstein-Hilbert term with cosmological constant in the bulk and the Gibbons-Hawking-York term on the lateral boundary. $R_g$ denotes the $g$-scalar curvature (following the signs conventions of \cite{Waldbook}), $\vol_g$ and $\vol_{\overline{g}}$ the metric volume forms, and $\xoverline{K}$ the extrinsic curvature of  $\partial_L\!M\subset M$ given by
\[\xoverline{K}:=\frac{1}{2}\jmath^*\!\L_{\vec{\nu}}g\qquad\equiv\qquad \xoverline{K}_{\overline{\alpha}\overline{\beta}}=\jmath^\alpha_{\overline{\alpha}}\jmath^\beta_{\overline{\beta}}\nabla_{\alpha}\nu_\beta\]
where $\vec{\nu}$ is the $g$-normal vector field to $\partial_L\!M$ with $\nu_\alpha\nu^\alpha=+1$.

\subsection*{Step 1 and 2: Variations}\label{subsection: variations metric}
According to the computations given in the appendix, we have
\[\dd L^{(m)}_{\mathrm{EH}}=E_{(m)}^{\alpha\beta}\wedge \dd g_{\alpha\beta}+\d\Theta_{(m)}\qquad\qquad \dd\xoverline{\ell}^{(m)}_{\mathrm{GHY}}-\jmath^*\Theta_{(m)}=\xoverline{b}_{(m)}^{\overline{\alpha}\overline{\beta}}\wedge\dd\xoverline{g}_{\overline{\alpha}\overline{\beta}}-\d\xoverline{\theta}_{(m)}\]
where
\begin{align*}
    &E_{(m)}^{\alpha\beta}(g)=-\left(\mathrm{Ric}^{\alpha\beta}-\left(\frac{R_g}{2}-\Lambda\right)g^{\alpha\beta}\right)\vol_g&&\xoverline{b}_{(m)}^{\overline{\alpha}\overline{\beta}}(g)=-\left(\xoverline{K}^{\overline{\alpha}\overline{\beta}}-\xoverline{K}\xoverline{g}^{\overline{\alpha}\overline{\beta}}\right)\vol_{\overline{g}}\\
    &\Theta_{(m)}(g)=\iota_{\vec{W}}\vol_g&&\xoverline{\theta}_{(m)}(g)=\iota_{\vec{V}}\vol_{\overline{g}}\\
    &W_{\!\alpha}=\nabla^\beta\dd g_{\alpha\beta}-\nabla_{\!\alpha}\dd g&&\xoverline{V}_{\!\overline{\alpha}}=-\jmath^\alpha_{\overline{\alpha}}(\iota_{\vec{\nu}}\dd g)_{\!\alpha}
\end{align*}
and $\dd g:=g^{\alpha\beta}\dd g_{\alpha\beta}$ stands for the $g$-trace of $\dd g_{\alpha\beta}$ (not to be confused with the variation of the determinant of $g$, which we do not use in this paper). Besides, we use the notation $(\iota_{\vec{\nu}}\dd g)_{\!\beta}:=\nu^\alpha\dd g_{\alpha\beta}$ even though $\dd g_{\alpha\beta}$ is not a $2$-form.

\subsection*{Step 3: Space of solutions}
Once we know the basic variations, we can compute the variation of the action 
\begin{equation}\label{eq: dd SS = 0}\dd_g\S\uptolids\int_M E_{(m)}^{\alpha\beta}(g)\wwedge \dd g_{\alpha\beta}-\int_{\partial_L\!M}\xoverline{b}_{(m)}^{\overline{\alpha}\overline{\beta}}(g)\wwedge\dd\xoverline{g}_{\overline{\alpha}\overline{\beta}}
\end{equation}
where $\uptolids$ denotes equality up to integrals over the lids. Notice that those integrals are unimportant because the values of the dynamical fields on the lids are kept fixed when obtaining the equations of motion of the theory. The solutions obtained from $\SS^{(m)}_N$ are given by metrics $g\in\F^{(m)}_N$ satisfying Einstein's equations $E_{(m)}(g)=0$ and the Neumann boundary condition $\xoverline{b}_{(m)}(g)=0$, whereas those from $\SS^{(m)}_D$ correspond to metrics $g\in\F^{(m)}_D$ satisfying Einstein's equation and no additional BC. Indeed, in the later case the boundary integral of \eqref{eq: dd SS = 0} vanishes as a consequence of the Dirichlet BC introduced \emph{a priori} on the definition of $\F^{(m)}_D$.

\begin{remark}\mbox{}\\
 It is important to realize that a boundary term must be included in the two theories that we consider (i.e. when no \emph{a priori} BC are imposed on the space of fields and with Dirichlet BC). Otherwise, as explained in section \ref{section: summary of algorithm}, the second step cannot be performed and thus the space of solutions is not well defined. Nonetheless, it is possible to impose homogeneous Neumann BC \emph{a priori} and include a different boundary term \cite{KrishnanRaju2017} (which vanishes for dimension 4). However, as we have shown, this is not necessary. Finally, it is worth noting that non-trivial BC may lead to constrained variations, which must be properly handled.
 \end{remark}
 
\subsection*{Step 4: Symplectic form}
Given a Cauchy embedding $\imath:\Sigma\hookrightarrow M$, we compute the presymplectic structure canonically associated with the action as
\[\OOmega_{(m)}^\imath=\dd\left(\int_\Sigma\imath^*\Theta_{(m)}-\int_{\partial\Sigma}\xoverline{\imath}^*\xoverline{\theta}_{(m)}\right)\]
Using the results of the previous sections and the computations of the appendix, we obtain:
\begin{equation}
\begin{split}
    \OOmega_{(m)}^\imath&=\frac{1}{2}\!\int_\Sigma n^\alpha\delta^{\lambda\eta\sigma}_{\alpha\beta\zeta}g^{\beta\rho}g^{\zeta\phi}\dd g_{\eta\rho}\wwedge\nabla_{\!\lambda}\dd g_{\sigma\phi}\vol_\gamma\\
&\phantom{=}-\frac{1}{2}\!\int_{\partial\Sigma}\!\Big(n^\alpha g^{\beta\eta}\mu^\lambda +m^\alpha\nu^\beta \xoverline{g}^{\overline{\eta}\overline{\lambda}}\jmath^\eta_{\overline{\eta}}\jmath^\lambda_{\overline{\lambda}}\Big)\dd g_{\alpha\beta}\wwedge\dd g_{\eta\lambda}\vol_{\overline{\gamma}}
\end{split}
\end{equation}
As indicated in \eqref{eq: diagrama}, $n^\alpha$ is the $g$-normal to $\imath(\Sigma)\subset M$, $\nu^\beta$ is the $g$-normal to $\partial_LM\subset M$, $m^\alpha$ is the pushforward of the $\xoverline{g}$-normal to $\xoverline{\imath}(\partial\Sigma)\subset\partial_L\!M$, and $\mu^\lambda$ is the pushforward of the $\gamma$-normal  to $\partial\Sigma\subset\Sigma$. We have also used the generalized Kronecker delta \[\tensor*{\delta}{*_{\alpha_1}^{\beta_1}_\cdots^\cdots_{\alpha_s}^{\beta_s}}=\begin{vmatrix}
\delta_{\alpha_1}^{\beta_1} & \cdots & \delta_{\alpha_1}^{\beta_s}\\ \vdots & &\vdots\\ \delta_{\alpha_s}^{\beta_1}&\cdots&\delta_{\alpha_s}^{\beta_s}\end{vmatrix}\]

\subsection*{Step 5: Charges}
The Noether currents associated with an arbitrary vector field $\xi\in\mathfrak{X}(M)$ tangent to the lateral boundary are computed in the appendix. The result is
\begin{align*}
    &J_\xi^{(m)}=\d Q_\xi^{(m)}+2\star_g(\iota_\xi\tilde{E}_{(m)})&&\xoverline{\jmath}_\xi^{(m)}=\jmath^* Q_\xi^{(m)}-2\star_{\overline{g}}(\iota_{\overline{\xi}}\tilde{b}_{(m)})
\end{align*}
where $Q_\xi^{(m)}=\star_g\d\xi$ is the $\xi$-Noether potential and $(\tilde{E}_{(m)},\tilde{b}_{(m)})$ are the $2$-tensors that multiply the top-forms in $(E_{(m)},\xoverline{b}_{(m)})$. As usual, $\star_g$ is the $g$-Hodge dual operator and, in order to ease the notation, we represent also by $\xi$ the $1$-form metrically equivalent to the vector field $\xi$.  The $\xi$-charge is then given by
\begin{equation}\label{eq: xi-charge}
\Q^\imath_{\xi,(m)}=2\int_\Sigma\imath^*\star_g\big(\iota_\xi\tilde{E}_{(m)}\big)+2\int_{\partial\Sigma}\xoverline{\imath}^*\star_{\overline{g}}\big(\iota_{\overline{\xi}}\tilde{b}_{(m)}\big)
\end{equation}
The last expression, when pulled back to $\Sol(\SS^{(m)}_N)\subset\F^{(m)}_N$, vanishes as expected because the theory is Diff-invariant \cite{MargalefVillasenor2020}. However, over $\Sol(\SS^{(m)}_D)\subset\F^{(m)}_D$, the boundary integral is not necessarily zero as $\xoverline{b}_{(m)}$ does not vanish in general. Notice that fixing the metric on the boundary breaks Diff-invariance.

\subsection*{Step 6: comparing with the Hamiltonian formulation}
If we consider $\partial\Sigma=\varnothing$, the equivalence between the CPS presymplectic form obtained in step 4 and the canonical ADM presymplectic form has been established in \cite{AshtekarMagnon1982,FrauendienerSparling1992}. The case with boundaries is not as straightforward and will be studied elsewhere.

\subsection*{The asymptotically flat case}
As a further application of the CPS-algorithm, let us obtain the ADM energy of an asymptotically flat space-time (see \cite{AshtekarBombelliReula1991,HarlowWu2020} for similar approaches). For that purpose, consider $M=\Sigma\times\R$ endowed with the Minkowski metric $g_0=-\d t^2+\sigma$. Then $\Sigma$ is a $3$-manifold with Euclidean metric $\sigma$ and boundary $\partial\Sigma=\S^2_r$, a $2$-sphere of $\sigma$-radius $r_0$. Take a small perturbation $h\in T_{g_0}\mathrm{Met}(M)$, such that $\xoverline{h}:=\jmath^*h=0$ (compatible with the Dirichlet condition, otherwise the charge is zero as the theory is Diff-invariant). The $\xi$-charge given by \eqref{eq: xi-charge} restricted to the space of solutions is given by
\[\Q_{\xi}=2\int_{\partial\Sigma}\xoverline{\xi}_{\overline{\alpha}}\xoverline{m}_{\overline{\beta}}(\xoverline{K}^{\overline{\alpha}\overline{\beta}}-\xoverline{K}\xoverline{g}^{\overline{\alpha}\overline{\beta}})\vol_{\overline{\gamma}}\]
Expanding the integrand in terms of $h$, and labeling with a circle the quantities associated with $g_0$ (the ``background'' objects), we obtain
\[\Q_{\xi}=2\int_{\partial\Sigma}\xoverline{\xi}_{\overline{\alpha}}(\accentset{\circ}{\xoverline{m}}_{\overline{\beta}}+\LL_h \xoverline{m}_{\overline{\beta}})\Big(\accentset{\circ}{\xoverline{K}}^{\overline{\alpha}\overline{\beta}}+\LL_h\xoverline{K}^{\overline{\alpha}\overline{\beta}}-(\accentset{\circ}{\xoverline{K}}+\LL_h \xoverline{K})(\accentset{\circ}{\xoverline{g}}^{\overline{\alpha}\overline{\beta}}-\LL_h\xoverline{g}^{\overline{\alpha}\overline{\beta}})\Big)(\vol_{\overline{\sigma}}+\LL_h\vol_\gamma)+\cdots
\]
where we omit higher order terms in $h$ as they vanish when $r_0\to\infty$ if we consider $\xi:=n$ and the inertial foliation $\{\{t\}\times\Sigma\}_t$ (in particular, we take $\vec{n}=\partial_t$).\vspace*{2ex}

Using the decomposition of the normal $\nu_\alpha=-\nu_\perp n_\alpha+\nu^\top_\alpha$ and some of the results of section \ref{subsection: some metric computations}, we have
\begin{align*}
\LL_h\xoverline{m}_{\overline{\alpha}}&=\jmath^\alpha_{\overline{\alpha}}\LL_h\!\left(\frac{n_\alpha+\nu_\perp \nu_\alpha}{|\nu^\top|}\right)=\ii_h\left(\frac{1}{|\nu^\top|}\jmath^\alpha_{\overline{\alpha}}\dd n_\alpha-\xoverline{m}_{\overline{\alpha}}\frac{\nu_\perp}{|\nu^\top|}\dd\nu_\perp\right)\\
&=\frac{1}{|\accentset{\circ}{\nu}^\top|}\jmath^\alpha_{\overline{\alpha}}\ii_h\left(\frac{-1}{2} n_\alpha\iota^2_{\vec{n}}\dd g\right)-\xoverline{m}_{\overline{\alpha}}\frac{\accentset{\circ}{\nu}_\perp}{|\accentset{\circ}{\nu}^\top|}\LL_h\nu_\perp\\
&=-\frac{1}{2} \xoverline{\xi}_{\overline{\alpha}}\iota^2_{\xi}\,h+0=-\frac{1}{2} \xoverline{\xi}_{\overline{\alpha}}\iota^2_{\overline{\xi}}\,\xoverline{h}=0
\end{align*}
where we have used that $\accentset{\circ}{\nu}_\perp=0$, $\xi=n$ and the fact that $\xi$ is tangent to the lateral boundary (hence we can replace $h$ with $\xoverline{h}$, which is zero). We have also used the (not very standard) notation $\iota^2_{V}T=V^\alpha V^\beta T_{\alpha\beta}$. Notice that the previous computation is performed at the point $g_0$ in the space of fields and at the lateral space-time boundary.\vspace*{2ex}

This computation together with  $\accentset{\circ}{\xoverline{m}}_{\overline{\beta}}=\xoverline{\xi}_{\overline{\beta}}$ takes care of the first parenthesis of $\QQ_\xi$. The last one can also be easily computed by using the variation of the volume given in section \ref{subsection: some metric computations}. Let us now deal with the term in the second parenthesis. First notice that $\accentset{\circ}{\xoverline{K}}^{\overline{\alpha}\overline{\beta}}$ vanishes when we contract it with $\xoverline{\xi}_{\overline{\alpha}}\xoverline{\xi}_{\overline{\beta}}$ (in coordinates this would be $\accentset{\circ}{\xoverline{K}}^{tt}$) while $\LL_h\xoverline{g}^{\overline{\alpha}\overline{\beta}}=\xoverline{h}=0$. A long but straightforward computation gives the following variations
\begin{align*}
\xoverline{\xi}_{\overline{\alpha}}\xoverline{\xi}_{\overline{\beta}}\LL_h\xoverline{K}^{\overline{\alpha}\overline{\beta}}&=\frac{1}{2}\xi^\alpha\xi^\beta\Big(\accentset{\circ}{\nabla}_{\!\vec{\nu}}h_{\alpha\beta}-2\nu^\gamma\accentset{\circ}{\nabla}_{\!\alpha}h_{\beta\gamma}\Big)\\
\LL_h\xoverline{K}&=\frac{1}{2}\imath^\alpha_a\imath^\beta_b\gamma^{ab}\Big(\accentset{\circ}{\nabla}_{\!\vec{\nu}}h_{\alpha\beta}-\nu^\gamma\accentset{\circ}{\nabla}_{\!\alpha}h_{\beta\gamma}\Big)-\frac{1}{2}\xi^\alpha\xi^\beta\Big(\accentset{\circ}{\nabla}_{\!\vec{\nu}}h_{\alpha\beta}-2\nu^\gamma\accentset{\circ}{\nabla}_{\!\alpha}h_{\beta\gamma}\Big)-\\
&\phantom{=}-\frac{1}{2}\accentset{\circ}{\xoverline{D}}_{\overline{a}}\Big(\xoverline{\gamma}^{\overline{a}\xoverline{b}}\xoverline{\jmath}^{\overline{\alpha}}_{\overline{b}}\jmath^\alpha_{\overline{\alpha}}(\iota_{\vec{\nu}}h)_\alpha\Big)+\frac{1}{2}\xi^\alpha\nu^\beta\L_{\vec{n}}g_{\alpha\beta}\xi^\gamma\nu^\mu h_{\gamma\mu}
\end{align*}
As the result is independent of the Cauchy slice, we introduce coordinates $\{t,x_1,x_2,x_3\}$ and take, as we mentioned before, a Cauchy slice given by $\Sigma=\{t=t_0\}$, then 
\[\vec{\nu}=\frac{1}{r_0}(0,x_1,x_2,x_3)\]
Notice that everything is constant along the $t$-direction (which is the $\xi$ direction because $\vec{n}=\partial_t$). Thus we have
\[ \accentset{\circ}{\xoverline{K}}_{\overline{\alpha}\overline{\beta}}=\jmath^\alpha_{\overline{\alpha}}\jmath^\beta_{\overline{\beta}}\accentset{\circ}{\nabla}_{\!\alpha}\nu_\beta\qquad\qquad\accentset{\circ}{\xoverline{K}}_{AB}=\frac{1}{r_0^3}(\delta_{AB}r_0^2-x_{A}x_{B})\qquad\qquad\accentset{\circ}{\xoverline{K}}=\frac{D-2}{r_0}\]
where here $A,B=1,2,3$ label the coordinates. Finally, notice that the integral of $h\accentset{\circ}{\xoverline{K}}\vol_{\overline{\sigma}}$ goes to zero in the limit $r_0\to\infty$ and $\iota_\xi \xoverline{K}=0$ because the normal is constant along the foliation. Putting everything together, we obtain
\[\Q_{\xi}=\int_{\partial\Sigma}\Big(\accentset{\circ}{\xoverline{K}}+\gamma^{BC}\nu^{A}\Big(\accentset{\circ}{\nabla}_{\!A}h_{BC}-\accentset{\circ}{\nabla}_{\!B}h_{CA}\Big)\Big)\vol_{\overline{\sigma}}+\cdots
\]
A final comment is now in order. As we can see we have a divergent constant term in the limit $r_0\to\infty$. To remove it, we introduce the following $r_0$-dependent boundary Lagrangian
\[\xoverline{\ell}_{(r_0)}=2\Big(\mathrm{Tr}_{\overline{g}}(\accentset{\circ}{\xoverline{K}})-\mathrm{Tr}_{\overline{g}}(\xoverline{K})\Big)\vol_{\overline{g}}\]
By doing this, the constant term does not appear, the limit is well defined, and we recover the well-known expression for the ADM-energy.

\section{General relativity in terms of tetrads}\label{section: tetrads}
\subsection*{Notation}
In order to introduce the tetrad formulation for GR, we will use internal abstract indices as discussed, for instance, in \cite{Penrose,AHM,Romano}. For the convenience of the unfamiliar reader, we have included this short subsection with the basic ingredients. The more familiar reader can go directly to equation \eqref{eq: ledge def}, where we introduce some new notation.\vspace*{2ex}

Let $\mathcal{V}\to M$ be a  Minkowski vector bundle i.e\ with typical fiber $V$\!, an $n$-dimensional vector space with internal indices $\{I,J,\ldots\}$, endowed with a Minkowskian metric $\eta$, which in abstract index notation reads $\eta_{IJ}$. This metric allows us to define the $\eta$-metric volume form $\varepsilon$, the $\eta$-Hodge dual $\star_\eta$, and the $\eta$-trace $\peqsub{\mathrm{Tr}}{\eta}$ (see \cite{Tecchiolli} for a recent review). The latter is given by
\[\peqsub{\mathrm{Tr}}{\eta}(\alpha\wedge\beta)=\eta^{I_1J_1}\cdots\eta^{I_1J_r}\alpha_{I_1\cdots I_r}\wedge \beta_{J_1\cdots J_r}\]
Tetrads are bundle isomorphisms $TM\rightarrow \mathcal{V}$. In abstract index notation they look like $e_{\alpha}^I$, so it is clear that a tetrad, together with the non-degenerate metrics $g_{\alpha\beta}$ and $\eta_{IJ}$, provides a way to swap space-time and internal indices. It is also clear from the index notation that tetrads can be identified with sections of the product bundle $\mathcal{V}\otimes T^*\!M$ i.e. elements of $\Gamma(\mathcal{V}\otimes T^*\!M)$, the set of $\mathcal{V}$-valued $1$-forms over $M$, although in practice it would be more useful to identify them with  $V$-valued $1$-forms over $M$. This is the case, for instance, if $M$ is parallelizable (an example would be a non-compact manifold admitting a spinor structure \cite{Geroch}). In that case it is enough to work with coframes which, loosely speaking, are the local version of a tetrad. Indeed, a coframe $e^I_\alpha(p)$ at $p\in M$ is a linear ismorphism $T_pM\rightarrow V$ satisfying
\[
g_{\alpha\beta}(p)=\eta_{IJ}e^I_\alpha(p)e^J_\beta(p)
\]
Equivalently, a coframe can be thought of as an orthonormal basis in $T^*_pM$. When $M$ is  parallelizable, coframes can be identified with global covector fields but, otherwise, they can only be considered as local covector fields, \cite{Romano}.\vspace*{2ex}

It is also useful to introduce objects with more indices, the so-called \textbf{generalized tensor fields} $t\in \Gamma(\mathcal{V}^{\otimes p} \otimes \mathcal{V}^{*\otimes q}\otimes TM^{\otimes r}\otimes T^*\!M^{\otimes s})$, where $\mathcal{V}^*\to M$ is the dual bundle of the  Minkowski bundle. In index notation, generalized tensor fields look like $t_{\alpha_1\cdots \alpha_s\, I_1\cdots I_q  }^{\beta_1\cdots \beta_p\,J_1\cdots J_r}$. We denote as $\Omega^k_r(M)$ the space of $k$-forms on $M$ with $r$ totally upper antisymmetric internal indices. Using Greek abstract indices $\{\alpha,\beta,\ldots\}$ for $M$ and capital Latin letters $\{I,J,\ldots\}$ for the internal indices, an element of $\Omega^k_r(M)$ will be written as \[\omega_{\alpha_1\cdots\alpha_k}^{I_1\cdots I_r}=\omega_{[\alpha_1\cdots\alpha_k]}^{[I_1\cdots I_r]}\]
To ease the notation, space-time indices will often be  omitted and we will simply write $\omega^{I_1\cdots I_r}$. Using the generalized Kronecker delta with internal indices, $\tensor*{\delta}{*_{I_1}^{J_1}_\cdots^\cdots_{I_m}^{J_m}}$, we define the graded wedge $\{\cdot{}\wedge\cdot{}\}:\Omega^k_r(M)\times\Omega^m_s(M)\to\Omega^{k+m}_{r+s}(M)$   
\begin{equation}
   \{A\wedge B\}_{I_1\cdots I_rJ_1\cdots J_s}:=\frac{1}{r!s!}\tensor*{\delta}{*_{I_1}^{K_1}_\cdots^\cdots_{I_r}^{K_r}_{J_1}^{L_1}_\cdots^\cdots _{J_s}^{L_s}}A_{K_1\cdots K_r}\wedge B_{L_1\cdots L_S}
\end{equation}
(this is analogous to the usual space-time wedge, although we have included curly brackets -- not necessary, strictly speaking -- as a reminder for those readers who may prefer explicit indices) and the graded bilinear product $[\cdot{}\ledge\cdot{}]:\Omega^k_r(M)\times\Omega^m_s(M)\to\Omega^{k+m}_{r+s-2}(M)$
\begin{equation}\label{eq: ledge def}[A\ledge B]_{I_1\dots I_{r-1}J_1\dots J_{s-1}}:=\tensor*{\delta}{*_{I_1}^{K_1}_{\dots}^{\dots}_{I_{r-1}}^{K_{r-1}}_{J_1}^{L_1}_{\dots}^{\dots}^{L_{s-1}}_{J_{s-1}}}    A_{K_1\dots K_{r-1}M}\wedge B^M_{\phantom{M}L_1\dots L_{s-1}}
\end{equation}
where the symbol $\,\ledge\,$ will be called the \emph{ledge} (Lie wedge). The previous operation is performed by contracting the last index of $A$ with the first one of $B$ and anti-symmetrizing the remaining ones. Although not necessary for our purposes, it is worth mentioning that $\Omega^k_2(M)$ with the ledge product
\[[A\ledge B]_{IJ}=A_{IK}\wedge \tensor{B}{^K_J}-A_{JK}\wedge \tensor{B}{^K_I}\]
 is a Lie algebra. Moreover, notice that $\{e\wedge e\}_{IJ}=2e^I\wedge e^J$. The following easy to derive formula will be used in the rest of the paper
\begin{align}
\label{eq: tr(a (b c))=-tr([a b] c)}
\peqsub{\mathrm{Tr}}{\eta}\big(\alpha\wedge\{\beta\wedge\gamma\}\big)=-\peqsub{\mathrm{Tr}}{\eta}\big([\alpha\ledge\beta]\wedge\gamma\big)\qquad\qquad\alpha\in\Omega^*_2(M),\ \ \beta,\gamma\in\Omega^*_1(M)
\end{align}
\subsection*{Constructing the geometric objects}
Consider $\{e_\alpha^I\}$ a co-frame, such that $\varepsilon_{I_1\cdots I_n}e^{I_1}\wedge\cdots\wedge e^{I_n}$ defines a volume form on $M$. We define the Lorentzian $e$-metric 
\[g=\peqsub{\mathrm{Tr}}{\eta}(e\otimes e)\qquad\equiv\qquad g_{\alpha\beta}=\eta_{IJ}e^I_\alpha e^J_\beta\]
together with its $g$-Levi-Civita connection $\nabla$. We can now define the dual frame $\{E_I^\alpha\}$, given by the relations
\[E_I^\alpha e^I_\beta=g^\alpha_\beta\qquad\qquad E_I^\alpha e^J_\alpha=\eta^J_I\]
as well as the $e$-connection $\omega$ defined by 
\[\nabla_{\!X} E_I=\tensor{\omega(X)}{^K_I}E_K\qquad\equiv\qquad\nabla_{\!\mu} E^\alpha_I=\omega_{\mu\,\,\,I}^{\,\,K} E^\alpha_K\qquad\equiv\qquad
\omega_{\mu\,\,\,I}^{\,\,K} =e_\alpha^K\nabla_{\!\mu}E^\alpha_I\]
which can be proven to uniquely determine $\omega$. Notice that the internal indices of $\omega$ are antisymmetric once the second one is raised $\omega_\mu^{IJ}=-\omega_\mu^{JI}$. Notice also that we have the covariant derivative of forms over $\Omega^r_s(M)$ given by
\[\mathcal{D}\alpha=\d\alpha+[\omega\ledge \alpha]\qquad\qquad \alpha\in\Omega^r_s(M)\]
where here and in the following, we will sometimes omit the internal indices when not needed. Let us also denote the associated $e$-Christoffel as $\Gamma$, the $e$-torsion as $T$, the $e$-curvature as $R$, and the $\omega$-curvature as $F$, which are given by
\begin{align*}
&\nabla_{\!E_I}E_J=\tensor{\Gamma}{^K_I_J}E_K&&\equiv&&e^I\tensor{\Gamma}{^K_I_J}=\tensor{\omega}{^K_J} 
\\
&\mathrm{Tor}(E_I,E_J)=\tensor{T}{^L_I_J}E_L&&\equiv&&\tensor{T}{^L_I_J}=e^L\Big(\mathrm{Tor}(E_I,E_J)\Big)\\
&\textrm{Riem}(E_I,E_J)E_K=\tensor{R}{^L_K_I_J}E_L&&\equiv&&\tensor{R}{^L_K_I_J}=e^L\Big(\textrm{Riem}(E_I,E_J)E_K\Big)\\
&F=\d\omega+\frac{1}{2}[\omega\ledge\omega]&&\equiv&&F_{IJ}=\d\omega_{IJ}+\omega_{IK}\wedge\tensor{\omega}{^K_J}
\end{align*}
where $\mathrm{Tor}$ and $\textrm{Riem}$ are the torsion and the Riemannian curvature of the $g$-Levi-Civita connection. In particular $\mathrm{Tor}=0$. It is interesting to notice that, as mentioned in the introduction of this section, we are using $E^\alpha_I$ and $e^I_\alpha$ to transform the space-time indices in $\mathrm{Tor}$ and $\textrm{Riem}$ into internal indices in order to define $T$ and $R$. From the previous equations, the following relations are easy to derive 
\begin{equation}\label{eq: F=R}
\tensor{F}{^I_J}=\frac{1}{2}\tensor{R}{^I_J_K_L}e^K\wedge e^L\qquad\qquad\qquad \tensor{T}{^L_I_J}=\mathcal{D}e^L(E_I,E_J)
\end{equation}
It is also straightforward to show that the following properties hold
\[(\mathcal{D}^2 \alpha)_{I_1\cdots I_r}=[F\ledge \alpha]_{I_1\cdots I_r}\qquad\qquad\qquad\qquad\mathcal{D}F_{IJ}=0\qquad\qquad\qquad\qquad\mathcal{D}e^I=0\]

\subsection*{Step 0: Action}
Let us consider the following spaces of non-degenerate tetrads on $M$
\begin{align*}
&\F^{(t)}_N=\{e\in\Omega^1_1(M)\ / \ e_Ie^I\in \F^{(m)}_N\}\\&\F^{(t)}_D=\left\{e\in\Omega^1_1(M)\ / \ e_Ie^I\in \F^{(m)}_N\text{ and } \xoverline{e}:=\jmath^*\!e\text{\; fixed}\right\}
\end{align*}
that we will refer to as Neumann and Dirichlet tetrads, respectively. The actions $\SS^{(t)}_N$ and $\SS^{(t)}_D$ are both defined by the same expression
\[\SS^{(t)}(e)=\int_M L^{(t)}(e)-\int_{\partial_LM}\xoverline{\ell}^{(t)}(e)\qquad\qquad\begin{array}{l}
L^{(t)}(e):=L^{(m)}_{\mathrm{EH}}\Big(\peqsub{\mathrm{Tr}}{\eta}(e\otimes e)\Big)
\\[1.2ex]\xoverline{\ell}^{(t)}(e):=\xoverline{\ell}^{(m)}_{\mathrm{GHY}}\Big(\peqsub{\mathrm{Tr}}{\eta}(e\otimes e)\Big)
\end{array}\] 
but suitably restricted to these spaces of tetrads. Notice that, essentially, we have only performed a ``change of variables'' from $g$ to $e$ through the (surjective but not injective) map
\[\Phi(e)_{\alpha\beta}=\eta_{IJ}e^I_\alpha e^J_\beta\]
Indeed, we have $\S^{(t)}=\Phi^*\S^{(m)}=\S^{(m)}\circ\Phi$. Since $\Phi$ is Lorentz invariant, the tetrad action is also Lorentz invariant i.e. if $\Psi\in SO(1,3)$, then $\SS(\Psi.e)=\SS(e)$. By using \eqref{eq: F=R}, it is possible to show that the Lagrangian admits the following explicit expression
\begin{equation}
L^{(t)}(e)=\frac{1}{2}\peqsub{\mathrm{Tr}}{\eta}\left(\!\Big(F-\frac{\Lambda}{12}\{e\wedge e\}\Big)\wedge\star_\eta\{e\wedge e\}\!\right)=\frac{1}{2}\varepsilon_{IJKL}\Big(F^{IJ}-\frac{\Lambda}{6}e^I \wedge e^J\Big)\wedge e^K\wedge e^L
\end{equation}
Finding a useful explicit expression for the boundary Lagrangian requires more work. Let us define the ``internal'' normal $N^I:=\iota_{\vec{\nu}}e^I\in\Omega^0_1(M)$, which satisfies $N_I\xoverline{e}^I_{\overline{\alpha}}=0$ and $N_IN^I=1$, and the ``internal'' projector $\gamma^I_J:=\eta^I_J-N^IN_J$. 
With these elements at hand, together with $\varepsilon_{IJKL}=\vol_{\alpha\beta\gamma\delta}E^\alpha_I E^\beta_J E^\gamma_K E^\delta_L$ and \eqref{eq: orientation sigma}, it is easy to rewrite $\xoverline{\ell}^{(t)}$ as
\[\xoverline{\ell}^{(t)}(e)=-\varepsilon_{IJKL}N^I\xoverline{\mathcal{D}}N^J\wedge\xoverline{e}^K\wedge \xoverline{e}^L\]
where
\begin{equation}\label{eq: overline D}
\xoverline{\mathcal{D}}A=\d A+[\xoverline{\omega}\ledge A]
\end{equation}
is the induced covariant derivative over the boundary given by the pullback connection $\xoverline{\omega}=\jmath^*\omega$. Recall also that $\xoverline{e}:=\jmath^*e$ which, in turn, allows us to define  its dual $\xoverline{E}_I^{\overline{\alpha}}:=\xoverline{g}^{\overline{\alpha}\overline{\beta}}\eta_{IJ}\xoverline{e}^J_{\overline{\beta}}$ which satisfies 
\[\xoverline{E}^{\overline{\alpha}}_I\xoverline{e}^I_{\overline{\beta}}=\xoverline{g}^{\overline{\alpha}}_{\overline{\beta}}\qquad\qquad\xoverline{E}^{\overline{\alpha}}_I\xoverline{e}^J_{\overline{\alpha}}=\gamma^K_I\]
With these ingredients, it is possible to obtain a better expression for $\xoverline{\ell}^{(t)}$ by considering
\begin{align*}
2&\varepsilon_{IJKL}\left(\!\left(N^I\d N^J{}-{}\frac{1}{2}\xoverline{\omega}^{IJ}\right)-N^I\xoverline{\mathcal{D}}N^J\right)\wedge\xoverline{e}^K\wedge \xoverline{e}^L\\
&=-\varepsilon_{IJKL}\left(\xoverline{\omega}^{IJ}+2N^I\xoverline{\omega}^{JK}N_K\right)\wedge\xoverline{e}^K\wedge \xoverline{e}^L=-\varepsilon_{IJKL}\xoverline{\omega}^{KJ}\left(\eta^I_K-2N^IN_K\right)\wedge\xoverline{e}^K\wedge \xoverline{e}^L\\
&=-\varepsilon_{IJKL}\xoverline{\omega}^{KA}(\gamma^J_A+N^JN_A)\left(\gamma^I_K-N^IN_K\right)\wedge\gamma^K_B\xoverline{e}^B\wedge \gamma^L_C\xoverline{e}^C\\
&=-\gamma^K_B\gamma^L_C\varepsilon_{IJKL}\xoverline{\omega}^{KA}\left(\gamma^J_A\gamma^I_K-(\gamma^J_AN^IN_K-N^JN_A\gamma^I_K)-N^JN_AN^IN_K\right)\wedge\xoverline{e}^B\wedge \xoverline{e}^C\\
&=-\gamma^K_B\gamma^L_C\gamma^J_A\gamma^I_K\varepsilon_{IJKL}\xoverline{\omega}^{KA}\wedge\xoverline{e}^B\wedge \xoverline{e}^C
\end{align*}
which is zero as all four internal indices of $\varepsilon$ are projected. This identity allows us to obtain, as already done for instance in \cite{BodendorferNeiman2013,CorichiRubalcavaVukasinac2016}, the alternative expression
\begin{equation}\label{eq: l(e)}
\xoverline{\ell}^{(t)}(e)=-\frac{1}{2}\peqsub{\mathrm{Tr}}{\eta}\Big(\big(\{N\wedge\d N\}-\overline{\omega}\big)\wedge \star_\eta\{\overline{e}\wedge\overline{e}\}\Big)=-\frac{1}{2}\varepsilon_{IJKL}(2N^I\d N^J-\overline{\omega}{}^{IJ})\wedge \overline{e}^K\wedge\overline{e}^L
\end{equation}

\subsection*{Step 1 and 2: Variations}\label{eq: b tetrad}
As before, we rely on the results and computations of the appendix to obtain
\[\dd L^{(t)}=E^{(t)}_I\wedge \dd e^I+\d\Theta^{(t)}\qquad\qquad \dd\xoverline{\ell}^{(t)}-\jmath^*\Theta^{(t)}=\xoverline{b}^{(t)}_I\wedge\dd\xoverline{e}^I-\d\xoverline{\theta}^{(t)}\]
where
\begin{align*}
&E^{(t)}_I(e)=-\big[\star_\eta F_{(\Lambda)}\ledge e\big]_I\\
&\xoverline{b}^{(t)}_I(e)=\big[\!\star_\eta(\{N\wedge\d N\}-\xoverline{\omega})\ledge\xoverline{e}\big]_I-2(\iota_{\overline{E}^J}\d\xoverline{e}^K)\star_\eta\!\{N\wedge\xoverline{e}\}_{JK}N_I\\
&\Theta^{(t)}=\frac{1}{2}\peqsub{\mathrm{Tr}}{\eta}\Big(\!\star_\eta\{e\wedge e\}\wedge\dd\omega\Big)\\
&\xoverline{\theta}^{(t)}=\frac{1}{2}\peqsub{\mathrm{Tr}}{\eta}\Big(\!\star_\eta\{\xoverline{e}\wedge\xoverline{e}\}\wedge\{N\wedge\dd N\}\Big)\\
&\displaystyle F_{(\Lambda)}^{IJ}:=F^{IJ}-\frac{\Lambda}{3}e^I\wedge e^J=\left(F-\frac{\Lambda}{6}\{e\wedge e\}\right)^{\!IJ}
\end{align*}

\subsection*{Step 3: Space of solutions}
The solutions derived from  $\SS^{(t)}_N$ are non-degenerate tetrads $e\in\F^{(t)}_N$ satisfying $E_I^{(t)}(e)=0$ and the ``Neumann'' boundary condition  $\xoverline{b}_I^{(t)}(e)=0$. The solutions obtained from $\SS^{(t)}_D$ are tetrads $e\in\F^{(t)}_D$ satisfying $E_I^{(t)}(e)=0$ and no additional condition at the boundary (the Dirichlet BC are part of the definition of $\F^{(t)}_D$).

\subsection*{Step 4: Symplectic form}
Given a Cauchy embedding $\imath:\Sigma\hookrightarrow M$, we have
\[\OOmega_{(t)}^\imath=\dd\!\left(\int_\Sigma\imath^*\Theta^{(t)}-\int_{\partial\Sigma}\xoverline{\imath}^*\xoverline{\theta}^{(t)}\right)\]
Using the results of the previous section, we immediately obtain
\begin{align*}
\OOmega_{(t)}^\imath=&\frac{1}{2}\int_\Sigma \peqsub{\mathrm{Tr}}{\eta}\Big(\!\star_\eta \dd\{e\wedge e\}\wwedge\dd\omega\Big)\\
&-\frac{1}{2}\int_{\partial\Sigma}\peqsub{\mathrm{Tr}}{\eta}\Big(\!\star_\eta\dd\{\xoverline{e}\wedge\xoverline{e}\}\wwedge\{N\wedge\dd N\}+\star_\eta\{\xoverline{e}\wedge\xoverline{e}\}\wedge\{\dd N\wwedge\dd N\}\Big)
\end{align*}

\subsection*{Step 5: Charges}
A direct computation using the definition of the $\xi$-current $J_\xi^{(t)}$ leads to
\[
J_\xi^{(t)}=\frac{1}{2}\peqsub{\mathrm{Tr}}{\eta}\Big(2[\star_\eta F_{(\Lambda)}\ledge e]\wedge\iota_\xi e-\star_\eta\mathcal{D}\{e\wedge e\}\wedge\iota_\xi\omega+\star_\eta\{e\wedge e\}\wedge(\L_\xi-\LL_{\XX_\xi})\omega\Big)-\d\peqsub{\mathrm{Tr}}{\eta}\Big(\frac{1}{2}\star_\eta\{e\wedge e\}\wedge\iota_\xi\omega\Big)
\]
where we have used $\iota_\xi F=(\L_\xi-\mathcal{D}\iota_\xi)\omega$, which follows from Cartan's magic formula. The second term of $J_\xi^{(t)}$ vanishes as the torsion $\mathcal{D}e$ is zero, while the third one is also zero as a consequence of \eqref{eq: definicion XX_xi} and the fact that the only background object is $\eta$, which is ``constant''. Thus we obtain
\begin{align*}
J_\xi^{(t)}=\peqsub{\mathrm{Tr}}{\eta}\Big(\iota_\xi e\wedge E^{(t)}\Big)-\d\peqsub{\mathrm{Tr}}{\eta}\Big(\frac{1}{2}\star_\eta\{e\wedge e\}\wedge\iota_\xi\omega\Big)\quad\longrightarrow\quad Q_\xi^{(t)}:=-\frac{1}{2}\peqsub{\mathrm{Tr}}{\eta}\Big(\!\star_\eta\{e\wedge e\}\wedge\iota_\xi\omega\Big)
\end{align*}

$J_\xi^{(t)}$ is then exact over the space of solutions with $\xi$-potential $Q_\xi^{(t)}$. Similarly, we have for the boundary
\begin{align*}
  \xoverline{\jmath}_\xi^{(t)}-\jmath^*Q_\xi^{(t)}&=-\frac{1}{2}\mathrm{Tr}\Big(\big\{N\wedge(\L_\xi-\LL_{\XX_\xi}\}N\big)\wedge\star_\eta\{\xoverline{e}\wedge\xoverline{e}\}-2\star_\eta(\{N\wedge\d N\}-\xoverline{\omega})\wedge\{\iota_{\overline{\xi}}\xoverline{e}\wedge\xoverline{e}\}\Big)\\
  &=\mathrm{Tr}\Big(0+[\star_\eta(\{N\wedge\d N\}-\xoverline{\omega})\ledge\xoverline{e}]\wedge\iota_{\overline{\xi}}\xoverline{e}\Big)\\
  &=\peqsub{\mathrm{Tr}}{\eta}(\iota_{\overline{\xi}}\xoverline{e}\wedge\xoverline{b}^{(t)})
\end{align*}

In the last equality, a term may seem to be missing according to the definition of $\xoverline{b}^{(t)}$ but, in fact, such term vanishes as a consequence of $\xoverline{e}^IN_I=0$. We then conclude that the $\xi$-charges are given by
\begin{equation}\label{eq: xi charge tetrad}\QQ^\imath_{\xi,(t)}=\int_\Sigma\imath^*\peqsub{\mathrm{Tr}}{\eta}\Big(\iota_\xi e\wedge E^{(t)}\Big)-\int_{\partial\Sigma}\xoverline{\imath}^*\peqsub{\mathrm{Tr}}{\eta}\Big(\iota_{\overline{\xi}}\xoverline{e}\wedge\xoverline{b}^{(t)}\Big)
\end{equation}
This expression, when pulled-back to $\Sol(\SS^{(t)}_N)\subset\F^{(t)}_N$, vanishes as expected because the theory is Diff-invariant \cite{MargalefVillasenor2020}. However, over $\Sol(\SS^{(t)}_D)\subset\F^{(t)}_D$ the boundary integral will not be zero in general, as $\xoverline{b}^{(t)}$ does not necessarily vanish (fixing the tetrad on the boundary breaks the invariance under diffeomorphisms).

\subsection*{Step 6: comparing with the Hamiltonian formulation}
From Step 6 of the CPS-algorithm in the metric formalism and the following section, the equivalence is assured when no boundaries are present. The case with boundaries is again not as straightforward and it will be studied elsewhere.

\section{Metric vs Tetrad formulation}\label{section: metricvstetrad}
We have obtained in the previous sections the presymplectic structure over the space of solutions for both the metric and tetrad formulations together with their $\xi$-charges. In this section, we prove that the spaces can be naturally mapped, that their symplectic structures are equivalent (without considering the internal gauge freedom), and that the $\xi$-charges are equal.

\subsection{Space of solutions}
Let us first show the correspondence between the solution spaces of metric and tetrad gravity. To this end, we define the maps
\[\Phi_N:\F^{(t)}_N\to\F^{(m)}_N\qquad\qquad\qquad\Phi_D:\F^{(t)}_D\to\F^{(m)}_D\]
both obtained by assigning the following metric to a given tetrad
\[\Phi(e)=\peqsub{\mathrm{Tr}}{\eta}(e\otimes e)\qquad\qquad\equiv\qquad\qquad\Phi(e)_{\alpha\beta}=\eta_{IJ}e^I_\alpha e^J_\beta\]
Of course, in the Dirichlet case the compatibility condition $\xoverline{g}_{\overline{\alpha}\overline{\beta}}=\eta_{IJ}\xoverline{e}^I_{\overline{\alpha}}\xoverline{e}^J_{\overline{\beta}}$ must hold for $\Phi_D$ to be well defined. On one hand, it is well known that $\Phi$ is surjective but not injective. For instance, $\Phi(-e)=\Phi(e)$. In fact, it can be proved that $\Phi(e)=\Phi(e')$ if and only if $e'_I=\tensor{\Psi}{_I^J}e_J$ for some $\Psi\in SO(1,3)$. \vspace*{2ex}

On the other hand, we have that $\S^{(t)}=\S^{(m)}\circ\Phi$ so
\[\dd_e\SS^{(t)}=\dd_e(\SS^{(m)}\circ\Phi)=\dd_{\Phi(e)}\SS^{(m)}\circ\dd_e\Phi\]

It is easy to check that $\dd_e\Phi$ is surjective. Hence, the relation between the spaces of solutions is clear: if $e\in\Sol(\SS^{(t)})$, then $\Phi(e)\in\Sol(\SS^{(m)})$ and if $g\in\Sol(\SS^{(t)})$, then every $e\in\Phi^{-1}(\{g\})$ belongs to $\Sol(\SS^{(t)})$.

\subsection{Presymplectic structures}
Let us now compare $\OOmega^\imath_{(m)}$ and $\OOmega^\imath_{(t)}$ by looking at their symplectic potentials. For that purpose, we perform the ``change of variable'' $g=\Phi(e):=e^Ie_I$ in $(\Theta_{(m)},\xoverline{\theta}_{(m)})$. First, we notice that instead of working with $\Theta_{(m)}$, it is more convenient to work with its $g$-dual
\begin{equation}\label{star Theta (m)}
(\star_g\Theta_{(m)})^\mu=(\star_g\iota_{\vec{W}}\vol_g)^\mu=W^\mu
\end{equation}
where $W_\mu:=\nabla^\beta\dd g_{\mu\beta}-\nabla_{\!\mu}\dd g$ was obtained in step 1  of section \ref{section: metric}. Using the variation
\[\dd\omega^{KL}_\mu=(\dd e^K_\alpha)\nabla_{\!\mu} E^{\alpha L}+E^{\alpha K}\nabla_{\!\mu}\dd e_\alpha^L-(e_{\alpha I}\dd e_\beta^I+e_{\beta I}\dd e_\alpha^I)E^{\alpha K}\nabla_{\!\mu}E^{\beta L}-E^{\beta K} e^L_\gamma(\dd\nabla)^\gamma_{\ \mu\beta}\]
which follows from the definition of  $\omega^{IJ}_\mu$, and the definition of $W^\mu$, we obtain on one hand
\begin{equation}\label{eq: W= cosas}
W^\mu=2E^\mu_I\iota_{E_J}\dd\omega^{IJ}+(\delta\mathcal{U})^\mu\qquad\qquad \mathcal{U}^{\alpha\mu}:=\big(g^{\beta\mu}E^\alpha_L-g^{\beta\alpha}E^\mu_L\big)\dd e^L_\beta
\end{equation}
where $(\delta\mathcal{U})^\mu:=-\nabla_{\!\alpha}\mathcal{U}^{\alpha\mu}$ is the codiferential. On the other hand, a standard computation using the definition of the $g$-Hodge star operator leads to
\begin{equation}\label{eq: star Theta (t)}
(\star_g\Theta^{(t)})^\mu=2E^\mu_I\iota_{E_J}\dd\omega^{IJ}
\end{equation}
Taking the Hodge dual of \eqref{eq: W= cosas}, using \eqref{star Theta (m)} and \eqref{eq: star Theta (t)}, and the fact that for $k$-forms we have $\star_g\star_g=(-1)^{k(n-k)+1}\mathrm{Id}$ and $\star_g\delta=(-1)^k\d\star_g$, we finally obtain
\[\Theta_{(m)}=\star_gW=\star_g\star_g\Theta^{(t)}+\star_g\delta\mathcal{U}=\Theta^{(t)}+\d\star_g\mathcal{U}\]

Once we have taken care of the bulk terms, we focus on the boundary ones. If we use  $\varepsilon_{IJKL}=\vol_{\alpha\beta\gamma\delta}E^\alpha_I E^\beta_J E^\gamma_K E^\delta_L$, $N_I=\nu_\alpha E^\alpha_I$, and $\dd N^I=-N_J\iota_{\overline{E}^I}\dd\xoverline{e}^J$, we can obtain on one hand
\begin{align*}
\big(\xoverline{\theta}^{(t)}\big)_{\overline{\alpha}\overline{\beta}}&=\tensor{\varepsilon}{^I^J_K_L}(\xoverline{e}^K\wedge\xoverline{e}^L)_{\overline{\alpha}\overline{\beta}}N_I\dd N_J\\
&=-\tensor{(\vol_g)}{^\alpha^\beta_\gamma_\delta}e^I_\alpha e^J_\beta E^\gamma_K E^\delta_L(\xoverline{e}^K_{\overline{\alpha}}\xoverline{e}^L_{\overline{\beta}}-\xoverline{e}^K_{\overline{\beta}}\xoverline{e}^L_{\overline{\alpha}})\nu_\sigma E^\sigma_IN_R\iota_{\overline{E}_J}\dd\xoverline{e}^R\\
&=-2N_R(\jmath^*\iota_{\vec{\nu}}\vol_g)_{\overline{\sigma}\,\overline{\alpha}\overline{\beta}}\xoverline{g}^{\overline{\sigma}\,\overline{\mu}}\dd\xoverline{e}^R_{\overline{\mu}}=(\iota_{\vec{U}}\vol_{\overline{g}})_{\overline{\alpha}\overline{\beta}}
\end{align*}
where $\xoverline{U}^{\overline{\mu}}:=-2N_R\xoverline{g}^{\overline{\mu}\,\overline{\alpha}}\dd\xoverline{e}^R_{\overline{\alpha}}$. On the other hand, using \eqref{eq: star 3+1}, we compute the following expression
\[\xoverline{\theta}_{(m)}-\xoverline{\theta}^{(t)}-\jmath^*(\star_g\mathcal{U})=\iota_{\vec{V}}\vol_{\overline{g}}-\iota_{\vec{U}}\vol_{\overline{g}}-\star_{\overline{g}}\jmath^*\iota_{\vec{\nu}}\mathcal{U}=\iota\big(\vec{V}-\vec{U}-\jmath^*\iota_{\vec{\nu}}\mathcal{U}\big)\vol_g\]
where $\xoverline{V}_{\overline{\alpha}}:=-\jmath^\alpha_{\overline{\alpha}}(\iota_{\vec{\nu}}\dd g)_{\!\alpha}$ was obtained in the step 2 of section \ref{section: metric}. Notice that we have made a small abuse of notation because $\jmath^*\iota_{\vec{\nu}}\mathcal{U}$ is a $1$-form so an index must be raised with the help of  $\xoverline{g}$. Finally, we show that this last expression is in fact zero
\begin{align*}\xoverline{V}^{\overline{\alpha}}&-\xoverline{U}^{\overline{\alpha}}-\xoverline{g}^{\overline{\alpha}\overline{\beta}}\jmath^\beta_{\overline{\beta}}g_{\mu\beta}\nu_\alpha\mathcal{U}^{\alpha\mu}\\
&=-\xoverline{g}^{\overline{\alpha}\overline{\beta}}\jmath^\beta_{\overline{\beta}}\nu^\alpha\eta_{IJ}(e_\alpha^I\dd e_\beta^J+e_\beta^I\dd e_\alpha^J)+2N_R\xoverline{g}^{\overline{\alpha}\overline{\beta}}\dd\xoverline{e}^R_{\overline{\beta}}-\xoverline{g}^{\overline{\alpha}\overline{\beta}}\jmath^\beta_{\overline{\beta}}g_{\mu\beta}\nu_\alpha(g^{\sigma\mu}E^\alpha_L-g^{\sigma\alpha}E^\mu_L)\dd e^L_\sigma\\
&=-N_J\xoverline{g}^{\overline{\alpha}\overline{\beta}}\dd \xoverline{e}^J_{\overline{\beta}}-\xoverline{g}^{\overline{\alpha}\overline{\beta}}\eta_{IJ}\xoverline{e}^I_{\overline{\beta}}\nu^\alpha\dd e_\alpha^J+2N_R\xoverline{g}^{\overline{\alpha}\overline{\beta}}\dd\xoverline{e}^R_{\overline{\beta}}-\xoverline{g}^{\overline{\alpha}\overline{\beta}}N_L\dd \xoverline{e}^L_{\overline{\beta}}+\xoverline{g}^{\overline{\alpha}\overline{\beta}}\nu^\sigma \eta_{LK}\xoverline{e}_{\overline{\beta}}^K \dd e^L_\sigma\\
&=0
\end{align*}
which proves that
\begin{equation}\xoverline{\theta}_{(m)}=\xoverline{\theta}^{(t)}+\jmath^*(\star_g\mathcal{U})
\end{equation}
Putting everything together, we finally obtain the main result of the paper
\begin{equation}\label{eq: (Theta,theta)=(Theta,theta)+d(A,a)}
	\comentario\tikzmarkin{potentialsesprima}(0.2,-0.35)(-0.2,0.55)
	\tikzmarkin{potentials}(0.2,-0.3)(-0.2,0.5)
	\Big(\Theta_{(m)},\xoverline{\theta}_{(m)}\Big)=\Big(\Theta^{(t)},\xoverline{\theta}^{(t)}\Big)+\underline{\d}\big(\!\star_g\mathcal{U},0\big)
	\tikzmarkend{potentials}
	\tikzmarkend{potentialsesprima}
\end{equation}
Hence, the symplectic potentials are equal up to a \emph{relative exact form}. Notice that the fact that the LHS comes from the metric formalism while the RHS comes from the tetrad formalism is not a problem because the LHS is implicitly evaluated at the $e$-dependent metric $\Phi(e):=\eta_{IJ}e^Ie^J$.\vspace*{2ex}

Taking now the $\dd$-exterior derivative of \eqref{eq: (Theta,theta)=(Theta,theta)+d(A,a)}, integrating over a Cauchy slice $(\Sigma,\partial\Sigma)$, and using the relative Stokes' theorem \eqref{eq: relative stokes}, we obtain the desired equality of the two presymplectic forms $\OOmega_{(t)}$ and $\OOmega_{(m)}$ (see appendix \ref{subsection: appendix mathematical background} for a brief account of the relative framework). More specifically, we have
\begin{equation}
\comentario\tikzmarkin{equivOmegaprima}(0.2,-0.26)(-0.2,0.41)
	\tikzmarkin{equivOmega}(0.2,-0.21)(-0.2,0.36)
    \OOmega_{(t)}=\Phi^*\OOmega_{(m)}
	\tikzmarkend{equivOmega}
	\tikzmarkend{equivOmegaprima}
\end{equation}
A final comment is in order now: the previous formula says that both presymplectic structures are equivalent \emph{modulo the gauge freedom} given by the kernel of $\Phi_*$. This gauge freedom is present in the tetrad formalism and originates in the $SO(1,3)$-invariance of $\Phi$ but it has no metric counterpart. In particular, this means that $\OOmega_{(t)}$ has more degenerate directions than $\OOmega_{(m)}$. This can be neatly understood by noticing that if we consider a curve $\Psi_\tau\in SO(1,3)$ and its associated vector $\mathbb{V}_{(t)}:=\left.\frac{\d}{\d\tau}\right|_0\Psi_\tau\cdot{} e\in T_e\F^{(t)}$, we have
\[\Psi_*\!\mathbb{V}_{(t)}=\left.\frac{\d}{\d\tau}\right|_0\Phi(\Psi_\tau\cdot{} e)=\left.\frac{\d}{\d\tau}\right|_0\Phi(e)=0\ \ \to\  \ii_{\mathbb{V}_{(t)}}\OOmega_{(t)}=\ii_{\mathbb{V}_{(t)}}\Phi^*\OOmega_{(m)}=\Phi^*\Big(\ii_{\Phi_*\!\mathbb{V}_{(t)}}\OOmega_{(m)}\Big)=0\]
$\mathbb{V}_{(t)}$ is a nonzero vector which belongs to the kernel of $\OOmega_{(t)}$ (as a consequence of being in the kernel of $\Psi_*$) so it is a gauge vector field. However, the metric counterpart $\mathbb{V}_{(m)}:=\Psi_*\!\mathbb{V}=0$ is not gauge because by definition the zero vector is not gauge. Finally, the equivalence of the charges \eqref{eq: xi-charge} and \eqref{eq: xi charge tetrad} is obvious as both theories are equivalent, implying that $\xoverline{b}_{(m)}$ and $\xoverline{b}^{(t)}$ are equivalent as well.

\section{Conclusions}

In this paper, we have studied the metric and tetrad formulations for general relativity on a manifold with boundary. By considering the appropriate bulk and boundary Lagrangians, we have shown that both theories are equivalent and hence, as one would expect, they are symplectically equivalent in the covariant phase space. Here we have focused on Dirichlet and Neumann BC, but any other BC will give the same results as long as the metric and tetrad actions are in a suitable correspondence.\vspace*{2ex}

It has been known for some time that, in the absence of boundaries, the metric symplectic current $\Omega_{(m)}:=\dd\Theta_{(m)}$ is equal to the tetrad symplectic current $\Omega^{(t)}:=\dd\Theta^{(t)}$ up to an exact form $\d A$ (and thus cohomologically equal). Therefore, their presymplectic forms over the space of solutions are equivalent since the integral of $\d A$ over a Cauchy slice $\Sigma$ is zero according to Stokes' theorem. On the other hand, if one considers a space-time with boundaries, the boundary-free covariant phase space procedure fails. This is due to some ambiguities that arise in the construction of the presymplectic form that hinder the direct comparison between metric and tetrad formulations. This has caused some discrepancies in previous works \cite{PaoliSpeziale2018,FGP1,JacobsonMohd2015}, but as we have shown, these difficulties arise because the traditional covariant phase space methods are only suited for the boundary-free case.\vspace*{2ex}

When boundaries are present, it is necessary to use more sophisticated techniques, like the relative bicomplex framework \cite{MargalefVillasenor2020}. Following the ideas of that formalism, we obtain the main result of the paper: the metric symplectic currents $(\Omega_{(m)},\xoverline{\omega}_{(m)}):=(\dd\Theta_{(m)},\dd\xoverline{\theta}_{(m)})$ and the tetrad symplectic currents  $(\Omega^{(t)},\xoverline{\omega}^{(t)}):=(\dd\Theta^{(t)},\dd\xoverline{\theta}^{(t)})$ are equal up to a \emph{relative exact form} $\underline{\d}(A,\xoverline{a})$. This implies, in particular, that they are equal in the relative cohomology (see appendix \ref{subsection: appendix mathematical background} for the relevant definitions) and that their presymplectic forms over the space of solutions are equivalent. Indeed, the relative Stokes' theorem tells us that the integral of $\underline{\d}(A,\xoverline{a})$ over a relative Cauchy slice $(\Sigma,\partial\Sigma)$ is zero. Furthermore, we have proved that the Noether charges are equivalent as expected. Finally, we have applied the covariant phase space methods to the asymptotically flat case to recover the well-known formula for the ADM-energy.

%%%%%%%%%%%%%%%%%%%%%%%%%%%%%%%%%%%%%%%%%%%%%%%%%%%%%%%%%%%%%%%%%%%%%%%
%
% ACKNOWLEDGMENTS
%
\section*{Acknowledgments}
The authors wish to thank Abhay Ashtekar, Laurent Freidel, and Simone Speziale for correspondence that prompted us to clarify some of the points discussed in the paper. This work has been supported by the Spanish Ministerio de Ciencia Innovaci\'on y Uni\-ver\-si\-da\-des-Agencia Estatal de Investigaci\'on FIS2017-84440-C2-2-P grant. Juan Margalef-Bentabol is supported by the Eberly Research Funds of Penn State, by the NSF grant PHY-1806356 and by the Urania Stott fund of Pittsburgh foundation UN2017-92945. E.J.S. Villase\~nor is supported by the Madrid Government (Comunidad de Madrid-Spain) under the Multiannual Agreement with UC3M in the line of Excellence of University Professors (EPUC3M23), and in the context of the V PRICIT (Regional Programme of Research and Technological Innovation).

%
% APPENDIX
%
\appendix
\section{Ancillary material}\label{section: appendix}
\subsection{Mathematical background}\label{subsection: appendix mathematical background}
\subsubsection*{Relative bicomplex framework}
In this section we include a summary of the definitions and results of \cite{MargalefVillasenor2020}. Consider an $n$-dimensional manifold $M$ with boundary $\partial M$ (possibly empty) and a space of fields $\F$ defined on it (sections of a bundle $E \to M$). The geometric structure of $\F$ may be understood by studying the infinite jet space of $E$. However, it is also possible to deal with $\F$ as is customary in the physics literature: by considering it as an infinite dimensional manifold endowed with standard operations such as the exterior derivative $\dd$, the interior product $\ii$, or the Lie derivative $\LL$. Physical field theories are described in terms of locally constructed fields over the space $M\times\F$, a space consisting of \textit{points of} $M$ and \textit{fields over} $M$.\vspace*{2ex}

We define the relative pair $(M,N)$ with $N\subset M$ being a submanifold $N\stackrel{\jmath}{\hookrightarrow} M$ of codimension $1$ of $M$. In this paper, we will always assume $N\subset\partial M$. Then, we have that the relative boundary of the pair is defined as
\[\underline{\partial}(M,N):=(\partial M\setminus N,\partial N)\]
which satisfies $\underline{\partial}^2=0$ and $\underline{\partial}(M,\partial M)=\varnothing$. The space of relative forms and the generalizations of some familiar operators to the present case are defined as
\begin{align*}
&\Omega^k(M,N):=\Omega^k(M)\oplus\Omega^{k-1}(N)&&\underline{\d}(A,\xoverline{a}):=(\d A,\jmath^*A-\d\xoverline{a})\\
&\underline{\iota}_V(A,\xoverline{a}):=(\iota_VA,-\iota_{\overline{V}}\xoverline{a})&&\underline{\star}_g(A,\xoverline{a}):=(\star_gA,\star_{\overline{g}}\xoverline{a})\\
&\underline{f}^*(A,\xoverline{a})=\Big(f^*A,(f|_N)^*\xoverline{a}\Big)&&\underline{\dd}(A,\xoverline{a})=(\dd A,\dd \xoverline{a})
\end{align*}
where $\xoverline{V}:=V|_N$ has to be tangent to $N$. Notice that $\underline{\d}^2=0$, hence,  we can define the so called relative cohomology $H^k(M,N)$. Two classes $[(A_1,\xoverline{a}_1)],[(A_2,\xoverline{a}_2)]\in H^k(M,N)$ are equal if and only if there exists $(B,\xoverline{b})\in\Omega^{k-1}(M,N)$ such that $(A_1,\xoverline{a}_1)=(A_2,\xoverline{a}_2)+\underline{\d}(B,\xoverline{b})$. The integral of a relative top-form $(A,\xoverline{a})\in\Omega^n(M,N)$ over the relative pair $(M,N)$ is defined as
\begin{equation}\label{eq: relative integral}\int_{(M,N)}(A,\xoverline{a}):=\int_M A-\int_N \xoverline{a}
\end{equation}
We have the relative Stokes' theorem given by
\begin{equation}\label{eq: relative stokes}\int_{(M,N)}\underline{\d}(B,\xoverline{b})=\int_{\underline{\partial}(M,N)}\underline{\jmath}^*(B,\xoverline{b})
\end{equation}
which in turn implies that \eqref{eq: relative integral} for $N=\partial M$ is well defined on relative cohomology because $\underline{\partial}(M,\partial M)=\varnothing$. We introduce now the space of forms $\OOmega^{(r,s)}(M\times\F)$ of degree $r$ in $M$ (horizontal part) and $s$ in $\F$ (vertical part). Endowed with the wedge product $\,\wwedge\,$, this space becomes a bigraded algebra with two exterior derivatives: the horizontal $\d$, which increases $r$ in one unit, and the vertical $\dd$, increasing $s$ in one unit. The wedge product $\,\wwedge\,$ restricted to $(k,0)$-forms coincides with $\wedge$. We will often abuse notation and use the latter. If we replace $(M,\d)$ by the relative pair $((M,N),\underline{\d})$, we can define the \textbf{relative bicomplex}
\[
\OOmega^{(r,s)}_{\mathrm{loc}}\Big((M,N)\times\F\Big)=\OOmega_{\mathrm{loc}}^{(r,s)}(M\times\F)\oplus\OOmega_{\mathrm{loc}}^{(r-1,s)}(N\times\F)\]
where the $\mathrm{loc}$ subscript indicates that we only consider $(r,s)$-forms which are locally constructed i.e. a form $\alpha$ evaluated at $p$ only depends on $p$, $\phi(p)$, and finitely many of the derivatives of $\phi$ at $p$.

\subsubsection*{Lagrangians and actions}

 \begin{definition}\label{star}\mbox{}\\
	We define a \textbf{Lagrangian pair} as an element of
	\[\Lag(M):=\OOmega_{\mathrm{loc}}^{(n,0)}((M,\partial M)\times\F)\]
\end{definition}
Remember that
\[[(L_1,\xoverline{\ell}_1)]=[(L_2,\xoverline{\ell}_2)]\ \quad \equiv\ \quad(L_2,\xoverline{\ell}_2)=(L_1,\xoverline{\ell}_1)+\underline{\d}(Y,\xoverline{y})\ \quad \equiv\ \quad \begin{array}{l}L_2=L_1+\d Y\\\xoverline{\ell}_2=\xoverline{\ell}_1+\jmath^*Y-\d\xoverline{y}\end{array}\]

\begin{definition}\mbox{}\\
	A \textbf{local action} is a map $\SS:\F\to\R$ of the form
	\begin{equation}\label{eq: action}
	\SS(\phi)=\int_{(M,\partial M)}(L,\xoverline{\ell})(\phi)
	\end{equation}
	for some local Lagrangian pair $(L,\xoverline{\ell})\in\Lag(M)$.
\end{definition}

\begin{definition}\label{def: S-equiv}\mbox{}\\
 $(L_i,\xoverline{\ell}_i)\in\Lag(M)$ are $\boldsymbol{\int}$\!\textbf{-equivalent}, which we denote as $(L_1,\xoverline{\ell}_1)\equivint(L_2,\xoverline{\ell}_2)$, if for every $\phi\in\F$, we have
	\[\int_{(M,\partial M)}(L_1,\xoverline{\ell}_1)(\phi)=\int_{(M,\partial M)}(L_2,\xoverline{\ell}_2)(\phi)\]
\end{definition}

In this work, we have only considered contractible bundles, for which the $\int$\!-equivalence is the same as the cohomological equivalence (a proof was given in \cite{MargalefVillasenor2020}). Nevertheless, if the bundles are not contractible, it is still possible to keep track of the ambiguities that arise from the fact that there exist non-zero Lagrangians $[(L,\xoverline{\ell})]\neq0$ whose Euler-Lagrange equations and BC are zero.

\subsubsection*{Variations}
We assume that the action is defined in such a way that it is possible to find Euler-Lagrange equations and boundary equations $(E,\xoverline{b})$, and symplectic potentials $(\Theta,\xoverline{\theta})$, such that
\begin{equation}\label{eq: dL}
\dd L=E_r\wedge\dd\phi^r+\d\Theta\qquad\qquad\qquad\dd\xoverline{\ell}-\jmath^*\Theta=\xoverline{b}_r\wedge\dd\phi^r-\d\xoverline{\theta}
\end{equation}
where $r$ labels the fields of the theory $\phi=(\phi^1,\ldots,\phi^R)\in\F$. If this is not possible, the theory is ill-posed and we have to change the space of fields $\F$ and/or the action $\SS$. The symplectic potentials $(\Theta,\xoverline{\theta})$ are defined up to a relative exact form. The space of solutions is \[\Sol(\SS):=\{\phi\in\F\ /\ (E,\xoverline{b})(\phi)=0\}\]

\subsubsection*{Symplectic structure}
We define the symplectic currents as $(\Omega,\xoverline{\omega}):=\underline{\dd}(\Theta,\xoverline{\theta})$. The relevant object is the relative integral of the symplectic currents over a Cauchy embedding $\imath:(\Sigma,\partial\Sigma)\hookrightarrow(M,\partial_LM)$
\[
\OOmega_\SS^\imath:=\int_{(\Sigma,\partial\Sigma)}\underline{\dd}\,\underline{\imath}^*\big(\Theta,\xoverline{\theta}\big)\in\OOmega^2(\F)\]
It can be proved that the pull-back of $\OOmega_\SS^\imath$ to the space of solutions is independent of the Cauchy embedding, endowing $\Sol(\SS)$ with  a presymplectic structure canonically associated with $\SS$.

\subsubsection*{Currents and charges}
Given some vector field $\xi\in\mathfrak{X}(M)$, we define the $\xi$-currents and the $\xi$-charges as
\[
(J_\xi,\xoverline{\jmath}_\xi):=\underline{\iota}_\xi (L,\xoverline{\ell})-\underline{\ii}_{\XX_\xi}(\Theta,\xoverline{\theta})\qquad\qquad\qquad\QQ^{\imath}_{\xi}:=\int_{(\Sigma,\partial\Sigma)}\underline{ \imath}^*(J_\xi,\xoverline{\jmath}_\xi)\in\OOmega^0(\F) \] 
	The $\xi$-charges in general depend on the chosen Lagrangians and on the embedding. If we compare the $\xi$-charges associated with two embeddings we obtain the following flux law
		\[\QQ^{\imath_2}_\xi-\QQ^{\imath_1}_\xi=\int_{(N,\lateral N)}(E_r,\xoverline{b}_r)(\phi)\L_\xi\phi^r+\int_{(N,\lateral N)}(\underline{\L}_\xi-\underline{\LL}_{\XX_\xi})(L,\xoverline{\ell})\]	
where $N$ is the manifold bounded by the Cauchy slices $\imath_1(\Sigma)$ and $\imath_2(\Sigma)$. In general, the charge $\QQ^{\imath}_\xi$ is not the Hamiltonian of the vector field $\XX_\xi\in\campos(\F)$  because

\[
\dd\QQ^{\imath}_\xi=\ii_{\XX_\xi}\OOmega^\imath_\SS+\int_{(\Sigma,\partial\Sigma)}\underline{\imath}^*\Big(\underline{\imath}_\xi(E_r,\xoverline{b}_r)\wedge\dd\phi^r\Big)+\int_{(\Sigma,\partial\Sigma)}\underline{\imath}^*(\L_\xi-\LL_{\XX_\xi})(\Theta,\xoverline{\theta})\]

The $\xi$-charge is the Hamiltonian of $\XX_\xi$ over $(\Sol(\SS),\OOmega_\SS)$ if and only if the last integral vanishes.

\subsection{Some computations in the metric case}\label{subsection: some metric computations}
Let us start off with a list of some of the well-known variations of the relevant objects used in the metric formalism
\begin{align*}
&\bullet\ \tensor{(\dd\nabla)}{^\alpha_\beta_\gamma}=\frac{1}{2}g^{\alpha\mu}\Big(\nabla_{\!\beta}\dd g_{\mu\gamma}+\nabla_{\!\gamma}\dd g_{\beta\mu}-\nabla_{\!\mu}\dd g_{\beta\gamma}\Big)&&\bullet\ \dd(g^{-1})^{\alpha\beta}=-g^{\alpha\mu}g^{\beta\nu}\dd g_{\mu\nu} \\
&\bullet\ \dd \mathrm{Ric}_{\beta\gamma}=\nabla_{\!\alpha}\tensor{(\dd\nabla)}{^\alpha_\beta_\gamma}-\nabla_{\!\beta}\tensor{(\dd\nabla)}{^\alpha_\alpha_\gamma}&&\bullet\ \dd \nu_{\alpha}=\frac{\iota^2_{\vec{\nu}}\dd g}{2}\nu_\alpha \\
&\bullet\ \dd R=-\mathrm{Ric}^{\alpha\beta}\dd g_{\alpha\beta}+\nabla^\alpha\nabla^\beta\dd g_{\alpha\beta}-\nabla^\alpha\nabla_{\!\alpha}\dd g&&\bullet\ \dd\vol_g=\frac{\dd g}{2}\vol_g\\
&\bullet\ \dd K_{\overline{\alpha}\overline{\beta}}=\frac{1}{2}\Big(K_{\overline{\alpha}\overline{\beta}}(\iota^2_{\vec{\nu}}\dd g)+\jmath_{\overline{\alpha}}^\alpha\jmath_{\overline{\beta}}^\beta\big(\nabla_{\!\vec{\nu}}\dd g_{\alpha\beta}-\nu^\mu\nabla_{\!\alpha}\dd g_{\mu\beta}-\nu^\mu\nabla_{\!\beta}\dd g_{\mu\alpha}\big)\Big)\span\span\\
&\bullet\ \dd (\mathrm{Tr}_{\overline{g}} K)=\frac{1}{2}\Big(\nabla_{\!\vec{\nu}}\dd g-\nu^\alpha\nabla^\beta\dd g_{\alpha\beta}-K^{\overline{\alpha}\overline{\beta}}\dd\xoverline{g}_{\overline{\alpha}\overline{\beta}}-\xoverline{\nabla}^{\overline{\beta}}\big(\jmath_{\overline{\beta}}^\beta(\iota_{\vec{\nu}}\dd g)_{\!\beta}\big)\Big)\span\span
\end{align*}
where $\dd g:=g^{\alpha\beta}\dd g_{\alpha\beta}$ and $\iota^2_{\vec{\nu}}\dd g:=\nu^\alpha\nu^\beta\dd g_{\alpha\beta}$.

\subsubsection*{Variations}
With those variations and the Lagrangians $(L^{(m)}_{\mathrm{EH}},\xoverline{\ell}^{(m)}_{\mathrm{GHY}})=\Big((R-2\Lambda)\vol_g,-2K\vol_{\overline{g}}\Big)$, we have
\begin{align*}
\underline{\dd}(L^{(m)}_{\mathrm{EH}},\xoverline{\ell}^{(m)}_{\mathrm{GHY}})=\left(E_{(m)},\xoverline{b}_{(m)}\right)\wedge\dd g+\underline{\d}\left(\Theta_{(m)},\xoverline{\theta}_{(m)}\right)
\end{align*}
where
\begin{align*}
&E_{(m)}^{\alpha\beta}=\left(\left(\frac{R}{2}-\Lambda\right)g^{\alpha\beta}-\mathrm{Ric}^{\alpha\beta}\right)\vol_g&&\xoverline{b}_{(m)}^{\overline{\alpha}\overline{\beta}}=\left(K^{\overline{\alpha}\overline{\beta}}-K\xoverline{g}^{\overline{\alpha}\overline{\beta}}\right)\vol_{\overline{g}}\\
&\Theta_{(m)}=\iota_{\vec{W}}\vol_g&&\xoverline{\theta}_{(m)}=\iota_{\vec{V}}\vol_{\overline{g}}\\
&W^\alpha=(g^{\alpha\mu}g^{\beta\lambda}-g^{\alpha\lambda}g^{\beta\mu})\nabla_{\!\lambda}\dd g_{\beta\mu}&&\xoverline{V}^{\overline{\alpha}}=-\xoverline{g}^{\overline{\alpha}\overline{\beta}}g^{\alpha\lambda}\jmath^\beta_{\overline{\beta}}\nu_\lambda\dd g_{\alpha\beta}
\end{align*}
\subsubsection*{Symplectic form}
The symplectic currents are given by
\begin{align*}
\Omega_{(m)}&:=\dd\Theta_{(m)}=\iota_{\dd \vec{W}}\vol_g-\iota_{\vec{W}}\dd\vol_g=\iota\!\left(\dd\vec{W}-\vec{W}\wwedge\frac{\dd g}{2}\right)\vol_g\\
\xoverline{\omega}_{(m)}&:=\dd\xoverline{\theta}_{(m)}=\iota_{\dd \vec{V}}\vol_{\overline{g}}-\iota_{\vec{V}}\dd\vol_{\overline{g}}=\iota\!\left(\dd\vec{V}-\vec{V}\wwedge\frac{\dd {\xoverline{g}}}{2}\right)\vol_{\overline{g}}
\end{align*}
To ease the notation, here we are using the parenthesis for the interior product instead of a subscript. These terms can then be rewritten as follows
\begin{align*}
g_{\alpha\beta}\left(\dd W^\beta-W^\beta\wwedge\frac{\dd g}{2}\right)&=\frac{1}{2}\dd g_{\alpha\mu}\wwedge\nabla^\mu\dd g+\frac{1}{2}g^{\sigma\beta}g^{\mu\lambda}\dd g_{\sigma\mu}\wwedge\nabla_{\!\alpha}\dd g_{\beta\lambda}+\frac{1}{2}\dd g\wwedge\nabla^\beta\dd g_{\alpha\beta}
\\
&\phantom{=}-\frac{1}{2}\dd g\wwedge\nabla_{\!\alpha}\dd g-g^{\mu\lambda}\dd g_{\beta\mu}\wwedge\nabla^\beta\dd g_{\alpha\lambda}\\
&=-\frac{1}{2}\delta^{\lambda\eta\sigma}_{\alpha\beta\zeta}g^{\beta\rho}g^{\zeta\phi}\dd g_{\eta\rho}\wwedge\nabla_{\!\lambda}\dd g_{\sigma\phi}+\frac{1}{2}\nabla^\eta(g^{\beta\lambda}\dd g_{\alpha\beta}\wwedge\dd g_{\eta\lambda})\\
\dd \xoverline{V}^{\overline{\alpha}}-\xoverline{V}^{\overline{\alpha}}\wwedge\frac{\dd \xoverline{g}}{2}&=\xoverline{g}^{\overline{\alpha}\,\overline{\gamma}}\overline{g}^{\overline{\beta}\,\overline{\delta}}\jmath^\beta_{\overline{\beta}}\dd\xoverline{g}_{\overline{\gamma}\overline{\delta}}\wwedge(\iota_{\vec{\nu}}\dd g)_{\!\beta}+\jmath^\beta_{\overline{\beta}}\xoverline{g}^{\overline{\alpha}\overline{\beta}}\Big\{g^{\alpha\gamma}(\iota_{\vec{\nu}}\dd g)_{\!\gamma}\wwedge\dd g_{\alpha\beta}\\
&\phantom{=}-\frac{1}{2}\iota^2_{\vec{\nu}}\dd g\wwedge(\iota_{\vec{\nu}}\dd g)_{\!\beta}+\frac{1}{2}(\iota_{\vec{\nu}}\dd g)_{\!\beta}\wwedge\dd g\Big\}\\
&=\frac{1}{2}\xoverline{g}^{\overline{\alpha}\overline{\beta}}\jmath^\beta_{\overline{\beta}}(\iota_{\vec{\nu}}\dd g)_{\!\beta}\wwedge(g^{\lambda\sigma}-\nu^\lambda\nu^\sigma)\dd g_{\lambda\sigma}=\frac{1}{2}\xoverline{g}^{\overline{\alpha}\overline{\beta}}\jmath^\beta_{\overline{\beta}}(\iota_{\vec{\nu}}\dd g)_{\!\beta}\wwedge \dd\xoverline{g}
\end{align*}

Consider a Cauchy embedding $\imath:\Sigma\hookrightarrow M$, where we have the $g$-normal $n^\alpha$ to $\imath(\Sigma)\subset M$, the $\xoverline{g}$-normal $\xoverline{m}^{\overline{\alpha}}$ to $\imath(\partial\Sigma)\subset\lateral M$, the $\gamma$-normal $\mu^b$ to $\partial\Sigma\subset\Sigma$, and the induced metric $\gamma=\imath^*g$. We denote also $m^\alpha:=\jmath^\alpha_{\overline{\alpha}}\xoverline{m}^{\overline{\alpha}}$ and $\mu^\beta_b:=\imath^\beta_b\mu^b$. Integrating the symplectic current over $(\Sigma,\partial\Sigma)$ we obtain the presymplectic form
\begin{align*}
\OOmega_{(m)}^\imath&=\int_{(\Sigma,\partial\Sigma)}\underline{\imath}^*(\Omega_{(m)},\xoverline{\omega}_{(m)})\updown{\eqref{eq: orientation sigma}}{\eqref{eq: reverse orientation}}{=}\\
&=-\int_\Sigma n_\alpha\left(\dd W^\alpha-W^\alpha\wwedge\frac{\dd g}{2}\right)\vol_\gamma-\int_{\partial\Sigma}\xoverline{m}_{\overline{\alpha}}\left(\dd \xoverline{V}^{\overline{\alpha}}-\xoverline{V}^{\overline{\alpha}}\wwedge\frac{\dd \xoverline{g}}{2}\right)\vol_{\overline{\gamma}}\\
&=\frac{1}{2}\!\int_\Sigma n^\alpha\delta^{\lambda\eta\sigma}_{\alpha\beta\zeta}g^{\beta\rho}g^{\zeta\phi}\dd g_{\eta\rho}\wwedge\nabla_{\!\lambda}\dd g_{\sigma\phi}\vol_\gamma\\
&\phantom{=}-\frac{1}{2}\!\int_{\partial\Sigma}\!\Big(n^\alpha g^{\beta\eta}\mu^\lambda +m^\alpha\nu^\beta \xoverline{g}^{\overline{\eta}\overline{\lambda}}\jmath^\eta_{\overline{\eta}}\jmath^\lambda_{\overline{\lambda}}\Big)\dd g_{\alpha\beta}\wwedge\dd g_{\eta\lambda}\vol_{\overline{\gamma}}
\end{align*}
Notice that we have used Stokes' theorem, Gauss' lemma (to write the covariant derivative $\nabla$ of $M$ in terms of the covariant derivative $D$ of $\Sigma$ and its extrinsic curvature) together with the fact that $T_{\alpha\eta}:=g^{\beta\lambda}\dd g_{\alpha\beta}\wwedge\dd g_{\eta\lambda}$ is antisymmetric (which kills the extrinsic curvature terms) to take $n_\alpha\nabla_{\!\eta} T^{\alpha\eta}$ to the boundary.

\subsubsection*{Charges}
The $\xi$-currents are given by
\begin{align*}
  \left(J^{(m)}_\xi,\xoverline{\jmath}^{(m)}_\xi\right)&=\underline{\iota}_\xi(L^{(m)}_{\mathrm{EH}},\xoverline{\ell}^{(m)}_{\mathrm{GHY}})-\underline{\ii}_{\XX_\xi}(\Theta_{(m)},\xoverline{\theta}_{(m)})=(\iota_{\vec{\xi}}L^{(m)}_{\mathrm{EH}},\iota_{\vec{\xi}}\,\xoverline{\ell}^{(m)}_{\mathrm{GHY}})-\Big(\iota_{\vec{W}}\vol_g,\iota_{\vec{V}}\vol_{\overline{g}}\Big)\\
  	&=\Big((R-2\Lambda)\iota_{\vec{\xi}}\vol_g,2K\iota_{\vec{\xi}}\vol_{\overline{g}}\Big)-\Big(\iota_{\underline{\ii}_{\XX_\xi}\vec{W}}\vol_g,\iota_{\underline{\ii}_{\XX_\xi}\vec{V}}\vol_{\overline{g}}\Big)\\
  	&=\Big(\iota\Big\{(R-2\Lambda)\vec{\xi}-\ii_{\XX_\xi}\vec{W}\Big\}\vol_g,\iota\Big\{2K\vec{\xi}-\ii_{\XX_\xi}\vec{V}\Big\}\vol_{\overline{g}}\Big)\\
  	&=\Big(\!\star_g\Big\{(R-2\Lambda)\xi-\ii_{\XX_\xi}W\Big\},\star_{\overline{g}}\Big\{2K\xoverline{\xi}-\ii_{\XX_\xi}\xoverline{V}\Big\}\Big)
\end{align*}
Using the definition of the Ricci and Riemann tensor together with \eqref{eq: definicion XX_xi}, we obtain
\[(R-2\Lambda)\xi_\alpha-\ii_{\XX_\xi}W_\alpha=(\delta\d\xi)_\alpha+2(\iota_{\vec{\xi}}\tilde{E})_\alpha\]
where $\delta$ is the codifferential (which is equal to minus the divergence) and $\tilde{E}_{(m)}^{\alpha\beta}:=E_{(m)}^{\alpha\beta}/\vol_g$ (the prefactor multiplying the volume form in $E_{(m)}$). Meanwhile, at the boundary 
\begin{align*}
2K\xoverline{\xi}^{\overline{\alpha}}-\ii_{\XX_\xi}\xoverline{V}^{\overline{\alpha}}&=2K\xoverline{\xi}^{\overline{\alpha}}+\xoverline{g}^{\overline{\alpha}\overline{\beta}}\jmath^\beta_{\overline{\beta}}\nu^\alpha(\nabla_{\!\alpha}\xi_\beta+\nabla_{\!\beta}\xi_\alpha)\\
&\!\!\overset{\vec{\xi}\perp\vec{\nu}}{=}2K\xoverline{\xi}^{\overline{\alpha}}+\xoverline{g}^{\overline{\alpha}\overline{\beta}}\jmath^\beta_{\overline{\beta}}\Big(\nu^\alpha(\d\xi)_{\alpha\beta}-2\xi_\alpha\nabla_{\!\beta}\nu^\alpha\Big)\\
&=2K\xoverline{\xi}^{\overline{\alpha}}+\xoverline{g}^{\overline{\alpha}\overline{\beta}}\Big(\jmath^\beta_{\overline{\beta}}\nu^\alpha(\d\xi)_{\alpha\beta}-2\xoverline{\xi}^{\overline{\mu}}K_{\overline{\mu}\overline{\beta}}\Big)\\
&=-2(K^{\overline{\alpha}\,\overline{\mu}}-\xoverline{g}^{\overline{\mu}\,\overline{\alpha}}K)\xoverline{\xi}_{\overline{\mu}}+\xoverline{g}^{\overline{\alpha}\overline{\beta}}\jmath^\beta_{\overline{\beta}}(\iota_{\vec{\nu}}\d\xi)_{\beta}\\
&=-2(\iota_{\vec{\xi}}\,\tilde{b})^{\overline{\alpha}}+(\jmath^*\iota_{\vec{\nu}}\d\xi)^{\overline{\alpha}}
\end{align*}
where $\tilde{b}:=b/\vol_{\overline{g}}$. Thus, using that for $2$-forms the equality $\delta\star_g=\star_g\d$ holds, we can write
\begin{align*}
\left(J^{(m)}_\xi,\xoverline{\jmath}^{(m)}_\xi\right)&=\Big(\!\star_g\delta\d\xi+2\star_g\iota_{\vec{\xi}}\tilde{E},-2\star_{\overline{g}}\iota_{\vec{\xi}}\,\tilde{b}+\jmath^*\#\iota_{\vec{\nu}}\d\xi\Big)\\
&=2\underline{\star}_g\underline{\iota}(\tilde{E},\tilde{b})+\Big(\d\star_g\d \xi,\jmath^*(\nu\wedge\#\d\xi+\#\iota_{\vec{\nu}}\d\xi)\Big)=2\underline{\star}_g\underline{\iota}(\tilde{E},\tilde{b})+\underline{\d}(\star_g\d \xi,0)
\end{align*}
where in the first line we have used the operator $\#\alpha=\star_g(\nu\wedge\alpha)$ which satisfies 
\begin{equation}\label{eq: star 3+1}
\star_g\alpha=\nu\wedge\#\alpha+\#\iota_{\vec{\nu}}\alpha\qquad\text{and}\qquad\jmath^*\#=\star_{\overline{g}}\jmath^*
\end{equation}

\subsection{Some tetrad computations}

\subsubsection*{Variations}

\begin{align*}
\dd L^{(t)}&=\frac{1}{2}\dd\peqsub{\mathrm{Tr}}{\eta}\left(\star_\eta\{e\wedge e\}\wedge\left(F-\frac{\Lambda}{12}\{e\wedge e\}\right)\right)=\\
&=\frac{1}{2}\peqsub{\mathrm{Tr}}{\eta}\left(2\star_\eta\{\dd e\wedge e\}\wedge F+\star_\eta\{e\wedge e\}\wedge \dd F-\frac{\Lambda}{3}\star_\eta\{\dd e\wedge e\}\wedge \{e\wedge e\}\right)=\\
&=\frac{1}{2}\peqsub{\mathrm{Tr}}{\eta}\left(2\star_\eta F\wedge \{e\wedge \dd e\}+\star_\eta\{e\wedge e\}\wedge \mathcal{D}\dd \omega-\frac{\Lambda}{3}\star_\eta\{e\wedge e\}\wedge \{e\wedge \dd e\}\right)\overset{\eqref{eq: tr(a (b c))=-tr([a b] c)}}{=}\\
&=\peqsub{\mathrm{Tr}}{\eta}\left(-\big[\!\star_\eta F_{(\Lambda)}\ledge e\big]\wedge \dd e-\frac{1}{2}\star_\eta\mathcal{D}\{e\wedge e\}\wedge \dd \omega\right)+\d\peqsub{\mathrm{Tr}}{\eta}\Big(\frac{1}{2}\star_\eta\{e\wedge e\}\wedge\dd\omega\Big)
\end{align*}
\begin{align*}
\dd\xoverline{\ell}^{(t)}&-\jmath^*\Theta^{(t)}=-\frac{1}{2}\dd\peqsub{\mathrm{Tr}}{\eta}\Big(\big(\{N\wedge\d N\}-\xoverline{\omega}\big)\wedge \star_\eta\{\xoverline{e}\wedge\xoverline{e}\}\Big)-\jmath^*\peqsub{\mathrm{Tr}}{\eta}\Big(\frac{1}{2}\star_\eta\{e\wedge e\}\wedge\dd\omega\Big)=\\
&=-\frac{1}{2}\peqsub{\mathrm{Tr}}{\eta}\Big(\big(\{\dd N\wedge\d N\}+\{N\wedge\d\dd N\}\big)\wedge \star_\eta\{\xoverline{e}\wedge\xoverline{e}\}+2(\{N\wedge\d N\}-\xoverline{\omega})\wedge \star_\eta\{\dd\xoverline{e}\wedge\xoverline{e}\}\Big)=\\
&=-\frac{1}{2}\peqsub{\mathrm{Tr}}{\eta}\Big(2\{\dd N\wedge\d N\}\wedge \star_\eta\{\xoverline{e}\wedge\xoverline{e}\}+\d\big(\{N\wedge\dd N\}\wedge \star_\eta\{\xoverline{e}\wedge\xoverline{e}\}\big)\\
&\qquad\qquad-2\{N\wedge \dd N\}\wedge \star_\eta\{\d\xoverline{e}\wedge\xoverline{e}\}+2\star_\eta(\{N\wedge\d N\}-\xoverline{\omega})\wedge \{\dd\xoverline{e}\wedge\xoverline{e}\}\Big)=\\
&=-\peqsub{\mathrm{Tr}}{\eta}\Big(\{\dd N\wedge\d N\}\wedge \star_\eta\{\xoverline{e}\wedge\xoverline{e}\}-\{N\wedge \dd N\}\wedge \star_\eta\{\d\xoverline{e}\wedge\xoverline{e}\}\\
&\qquad\qquad-[\star_\eta(\{N\wedge\d N\}-\xoverline{\omega})\ledge \xoverline{e}]\wedge\dd\xoverline{e}\Big)-\d\peqsub{\mathrm{Tr}}{\eta}\left(\frac{1}{2}\{N\wedge\dd N\}\wedge \star_\eta\{\xoverline{e}\wedge\xoverline{e}\}\right)
\end{align*}
Let us prove that the first term vanishes. First notice that $\varepsilon^{[IJKL}N^{M]}\in\Omega^0_5(M)=\{0\}$. Thus
\begin{align*}
    0&=5\varepsilon^{[IJKL}N^{M]}\xoverline{e}_K\wedge\xoverline{e}_L\wedge\mathrm{d}N_I\wedge(\iota_{\overline{E}_J}\dd \xoverline{e}_M) =\\
    &=\big(\varepsilon^{IJKL}N^{M}+\varepsilon^{JKLM}N^{I}+\varepsilon^{KLMI}N^{J}+2\varepsilon^{LMIJ}N^{K}\big) \xoverline{e}_K\wedge\xoverline{e}_L\wedge\mathrm{d}N_I\wedge(\iota_{\overline{E}_J}\dd \xoverline{e}_M)=\\
    &=-\varepsilon^{IJKL} \xoverline{e}_K\wedge\xoverline{e}_L\wedge\mathrm{d}N_I\wedge\dd N_J=-\star_\eta\{\xoverline{e}\wedge\xoverline{e}\}_{IJ}\wedge\d N^I\wedge\dd N^J
    \end{align*}
    where we have used $\dd N_I=-N_J\iota_{\overline{E}_I}\dd\xoverline{e}^J$, $N^I\xoverline{E}_I^{\overline{\alpha}}=0$, $N^I\mathrm{d}N_I=0$, and $N^I\xoverline{e}_I=0$. Let us now rework the second term to obtain the desired expression of page \pageref{eq: b tetrad}
\begin{align*}
	&\peqsub{\mathrm{Tr}}{\eta}\Big(\{N\wedge \dd N\}\wedge \star_\eta\{\d\xoverline{e}\wedge\xoverline{e}\}\Big)=\{N\wedge \dd N\}^{IJ} \frac{1}{2}\varepsilon_{IJKL}\{\d\xoverline{e}\wedge\xoverline{e}\}^{KL}\\
	&=2N^I\dd N^J\varepsilon_{IJKL}(\d\xoverline{e}^K)\wedge\xoverline{e}^L=-2N^IN_R(\iota_{\overline{E}^J}\dd\xoverline{e}^R)\varepsilon_{IJKL}(\d\xoverline{e}^K)\wedge\xoverline{e}^L\\
	&=-2N^IN_R\varepsilon_{IJKL}\left(\iota_{\overline{E}^J}\Big(\dd\xoverline{e}^R\wedge\d\xoverline{e}^K \wedge\xoverline{e}^L\Big)+\dd\xoverline{e}^R\wedge(\iota_{\overline{E}^J}\d\xoverline{e}^K)\wedge\xoverline{e}^L+\dd\xoverline{e}^R\wedge\d\xoverline{e}^K \wedge\gamma^{JL}\right)\\
	&=-2N^IN_R\varepsilon_{IJKL}\left(0+\dd\xoverline{e}^R\wedge(\iota_{\overline{E}^J}\d\xoverline{e}^K)\wedge\xoverline{e}^L+0\right)=-2N^IN_R\varepsilon_{IJKL}(\iota_{\overline{E}^J}\d\xoverline{e}^K)\wedge\xoverline{e}^L\wedge\dd\xoverline{e}^R\\
	&=-N_R\varepsilon_{IJKL}(\iota_{\overline{E}^J}\d\xoverline{e}^K)\{N\wedge\xoverline{e}\}^{IL}\wedge\dd\xoverline{e}^R=-2N_R(\iota_{\overline{E}^J}\d\xoverline{e}^K)\star_\eta\!\{N\wedge\xoverline{e}\}_{JK}\wedge\dd\xoverline{e}^R
\end{align*}

\end{document}